\documentclass[twocolumn, usenatbib]{mnras}
\usepackage{savesym}
\usepackage{graphicx}
\expandafter\let\csname equation*\endcsname\relax
  \expandafter\let\csname endequation*\endcsname\relax 
\usepackage{subfig}
\usepackage{amsmath}
\usepackage{amssymb}
\usepackage{verbatim}
\usepackage[yyyymmdd,hhmmss]{datetime}
\usepackage{array}
\usepackage{times}

\newcommand{\beq}{\begin{equation}}
\newcommand{\eeq}{\end{equation}}

\title[Random walks into black holes ]{ The dynamics of accretion flows near to the innermost stable circular orbit    }
\author [Andrew Mummery, Francesco Mori, Steven Balbus]{Andrew Mummery\thanks{E-mail:
andrew.mummery@physics.ox.ac.uk}, Francesco Mori, Steven Balbus
\\
Oxford Theoretical Physics, Beecroft Building,  Clarendon Laboratory, Parks Road, Oxford, OX1 3PU, United Kingdom}
\begin{document}

\pagerange{\pageref{firstpage}--\pageref{lastpage}} \pubyear{2022}
\date{}

\maketitle

\label{firstpage}

\begin{abstract} 
Accretion flows are fundamentally turbulent systems, yet are {classically modelled} with {viscous theories only valid on length scales significantly greater than the typical size of turbulent eddies in the flow}. We demonstrate that, while this will be a reasonable bulk description of the flow at large radii,  this must break down as the flow approaches absorbing boundaries, such as the innermost stable circular orbit (ISCO) of a black hole disc. This is because in a turbulent flow large velocity fluctuations can carry a fluid element over the ISCO from a finite distance away, from which it will not return,  a process without analogy in {conventional models}. This introduces a non-zero directional bias into the velocity fluctuations in the near-ISCO disc.  By studying {reduced} random walk problems, we derive a number of implications of the presence of an absorbing boundary in an accretion context.  In particular, we show that the average velocity with which a typical fluid element crosses the ISCO is much larger than is assumed in traditional theories. This enhanced velocity  modifies the thermodynamic properties of black hole accretion flows on both sides of the ISCO.  In particular, thermodynamic quantities for larger ISCO stresses no longer display pronounced cusps at the ISCO in this new formalism, a result with relevance for a number of observational probes of the intra-ISCO region. Finally, we demonstrate that these extended models reproduce the trans-ISCO behaviour observed in  GRMHD simulations of thin discs. 
\end{abstract}

\noindent

\section{Introduction} 
The accretion of material onto astrophysical black holes liberates vast amounts of energy and is the process through which some of the brightest objects in the Universe are powered. The modelling of these astrophysical accretion flows represents one of the original probes of the strong field regime of gravity, through which the properties of numerous black holes have been constrained \citep[e.g.,][]{Reynolds13, McClintock14}. One such prediction of general relativity, the existence of an innermost stable circular orbit \citep[hereafter ISCO;][]{Bardeen72}, profoundly modifies the dynamic and thermodynamic properties of accreting fluids at short distances from a black hole. Within this ISCO radius circular motion is unstable to inwards perturbations, and test particles plunge towards the singularity at $r=0$.  As a fundamental prediction of general relativity, the development of theoretical descriptions of the observational characteristics of this strong field regime  may be leveraged in the future to derive tighter constraints on black hole properties \citep[see e.g.,][for an example of this philosophy applied to iron line fitting]{Reynolds97, Wilkins20}.  For this to be a viable approach we must be sure that our theoretical descriptions of accretion accurately {reflect} the physical conditions of astrophysical sources. 

Accretion flows are fundamentally turbulent systems, a result of the magneto-rotational instability \citep[MRI;][]{BalbusHawley91}. However, all conventional analytical modelling of accretion flows assume that they can be described by an effective ``viscous'' redistribution of angular momentum \citep[e.g.,][]{SS73, NovikovThorne73}. {These classical} model{s} of {accretion flows must break down on length scales shorter than the typical size of a disc turbulent eddy, which for an accretion flow is macroscopic -- of order the disc's scale height $H$.}  When there is no {radial} length scale which {probes} the {  disc} fluctuation {scale} (i.e, far out in the main body of the disc {where observational diagnostics can be safely averaged over many disc scale heights}), the effects of {this  simplification are} likely inconsequential (which is why conventional thin disc theory works so well in this limits). However, as we argue here, once there is a relevant {radial} length scale with which to contrast {with the eddy scale} (such as the length scale associated with the inner disc edge), {this classical} prescription {should break down, and new descriptions should be developed which may well} result in different, and observationally relevant, predictions. 

In an attempt to move away from {these classical} descriptions, in this paper we examine the properties of a series of {reduced} random walk models. Systems undergoing a random walk are more mathematically flexible than purely viscous systems, and are well suited to modelling physical systems with both a global diffusive character (like an accretion flow {on} large {scales}), but also with large amplitude velocity fluctuations (like an accretion flow at {turbulent eddy scales}).  These random walk calculations highlight that the typical velocity with which a fluid element crosses the ISCO may be orders of magnitude higher than predicted from classical thin disc models with a finite ISCO stress \citep[e.g.,][]{AgolKrolik00}. It was recently demonstrated \citep{MummeryBalbus2023} that the trans-ISCO velocity plays a key role in the thermodynamic evolution of fluid flows inside of the ISCO, and this velocity amplification therefore produces important modifications to these profiles in the near-ISCO region. 

While we motivate this calculation on purely physical grounds, it could be equally well motivated from a purely model comparison perspective. Models of 
accretion with a large finite stress \citep[e.g.,][]{AgolKrolik00, MummeryBalbus2023} show pronounced cusps at the ISCO. These cusps{, or to be more precise discontinuities in the gradients of thermodynamic properties,} are almost certainly unphysical, and a result purely of the governing assumptions of {classical} accretion theory. Indeed, GRMHD simulations of accretion flows display a smooth evolution across the ISCO, with 
no indication of cusps \citep[e.g.,][]{Shafee08, Noble10, Zhu12, Schnittman16, Liska22}, even for those simulations which display large ISCO stresses. It seems likely that ISCO-cusps in analytical models of accretion could produce systematic effects when such models are fit to data \citep[e.g.,][]{Reynolds97, Wilkins20}, and therefore it is of general modelling interest to examine the physical causes of such behaviour, and improve on the underlying modelling assumptions in such instances. 

The layout of this paper is as follows. In section \ref{discussion} we review the classical physics of accretion relevant for this near-ISCO study. In section \ref{rwsettup} we introduce the mathematical random walk models we shall consider in this paper, before solving some explicit {reduced} problems in section \ref{explicit}. In section \ref{implications} we put these results in an astrophysical context, and derive modified global thin disc solutions which take into account the insight gained from the random walk models. We compare these models to some GRMHD simulations of thin discs in section \ref{grmhdcomp}, finding good agreement, before concluding in section \ref{conc}.

\section{ Near-ISCO accretion flows  }\label{discussion}
In this section we construct an argument based on classical disc theory which highlights that in the presence of absorbing boundaries {the classical} description of the mean fluid flow {of an accretion disc} must break down. 

The classical theory of {extra}-ISCO relativistic accretion proceeds by first  defining  a stress energy tensor which describes  the accretion flow and then by constructing mass, energy and momentum conservation equations. The classical accretion disc stress energy tensor is the following \citep[e.g.,][]{NovikovThorne73, Balbus17} 
\beq
T^{\mu\nu} = \left(\rho + {P + e \over c^2}\right) u^\mu u^\nu + P g^{\mu\nu} + {1\over c^2}(q^\mu u^\nu + q^\nu u^\mu) ,
\eeq
where $\rho$ is the rest mass density, $e$ the {internal} energy density and $P$ the pressure of the fluid. The 4-velocity of the flow is $u^\mu$, while $u_\mu$ is its covariant counterpart.  The final pair of terms represent the energy-momentum flux carried out of the system by photons, where $q^\mu$ is the photon flux 4-vector. 
 
With this stress energy tensor defined, one solves the equations of mass, angular momentum and energy conservation 
\beq
\nabla_\mu (\rho u^\mu) = 0 , \quad \nabla_\mu (T^\mu_\phi) = 0, \quad \nabla_\mu (T^\mu_0) = 0, 
\eeq
where in these expressions $\nabla_\mu$ is a covariant derivative with respect to Kerr metric coordinate $x^\mu$. 
These three constraints are sufficient to determine  three key  quantities: the governing equation for the evolution of the disc surface density, the radial velocity of the flow, and the energy flux out of the upper and lower disc surfaces \citep[see e.g.,][for discussions and various derivations]{NovikovThorne73, Balbus17}.   

The principal theoretical simplification employed in deriving the thin disc solutions of these coupled equations pertains to a series of approximations regarding the properties of the disc fluid's velocity. In particular,  the solutions to these three equations are derived by making the following important assumption: the total disc 4-velocity $u^\mu$ (as well as  $u_\mu$) may be decomposed into a mean component $U^\mu$ and vanishing-mean fluctuating component $\delta U^\mu$:
\beq
u^\mu = U^\mu + \delta U ^\mu ,\quad u_\mu = U_\mu + \delta U_\mu ,
\eeq
which satisfy asymptotic scalings \citep{BalbPap99, Balbus17}
\beq\label{main_scalings}
\delta U_\phi \ll U_\phi, ~~~ U^z \ll U^r \ll \delta U^r \sim \delta U_\phi / r \ll rU^\phi .
\eeq
While the fluctuations are an asymptotic scale larger than the mean radial flow of the disc, they are assumed to vanish on average 
\beq\label{fluc_int}
\left\langle \delta U^\mu \right\rangle \equiv {1 \over \Delta t} \int_t^{t+\Delta t} \delta U^{\mu}(r, t') \, {\rm d}t' = 0,
\eeq
where $\Delta t$ is a time long compared to the timescale upon which turbulent fluctuations are induced in the flow, but much shorter than the timescale upon which the mean disc quantities evolve. {Physically, this timescale should be thought of as a few times the orbital period at a radius $r$.} While the fluctuations themselves are assumed to vanish on average, their  average correlations are in general non-zero. In particular, accretion is ultimately driven by the non-zero correlation of the components of the turbulent velocity fluctuations, which produce a turbulent stress tensor $W^{\mu\nu}$:
\beq
W^{\mu\nu} \equiv \left\langle \delta U^\mu \delta U^\nu \right\rangle ,
\eeq 
where the angled brackets denote the same averaging procedure introduced above.  As the first order fluctuations in the disc velocity vanish on average, and the second order drift velocity is assumed to be extremely small, the zeroth order motion of the disc is well approximated by that of precisely circular motion, i.e.,  $U^0, U_0, U^\phi$ and $U_\phi$ are equal to the test particle circular motion solutions of the Kerr metric.

By making certain assumption about the local properties of $W^{\mu \nu}$, classical relativistic accretion models \citep[e.g., the][\citealt{PageThorne74}, and \citealt{SS73} solutions]{NovikovThorne73} allow the mean second-order radial accretion velocity of the flow at a given radius to be determined as a function of the physical parameters of the system (e.g., the black hole mass and spin, the mass accretion rate, and the disc stress $\alpha$-parameter).  However, it is important to recall that accretion flows are turbulent, and that the typical scale of the turbulent fluctuations are of a different (larger)  asymptotic scale than the mean drift velocity 
\beq
U^r \ll \delta U^r . 
\eeq
In the main bulk of the disc (far from the ISCO radius $r_I$), these velocity fluctuations vanish on average 
\beq
\left\langle \delta U^r(r\gg r_I) \right\rangle = {1 \over \Delta t} \int_t^{t+\Delta t} \delta U^r(r\gg r_I, t') \, {\rm d}t' = 0. 
\eeq
In effect this statement follows from the fact that there is no preferred direction in which the turbulent fluctuations occur, and fluctuations (e.g.) outwards in the disc are all compensated by turbulent fluctuations inwards. 
However, close to the ISCO itself, these velocity fluctuations will develop a non-zero directional bias, with fluctuations across the ISCO in effect absorbed into the black hole due to the lack of rotational support within the ISCO. This means that fluctuations across  the ISCO from the main body of the disc are no longer compensated for by fluctuations back from the intra-ISCO region. This favouring of fluctuations in a specific direction will mean that the average defined above will no longer vanish in the near-ISCO region, and instead
\begin{multline}
\left\langle \delta U^r(r\sim r_I) \right\rangle = {1 \over \Delta t} \int_t^{t+\Delta t} \delta U^r(r\sim r_I, t') \, {\rm d}t'  \\ \sim - \left| \delta U^r(r \sim r_I) \right| \gg U^r .
\end{multline}
Therefore, careful attention must be paid to the precise value of the the trans-ISCO velocity used in computing the evolution of the intra-ISCO thermodynamic quantities.

{ Of course, the ISCO does not represent a truly perfect absorbing boundary in a black hole accretion flow (only the event horizon is truly such a boundary). It is in principle possible for a fluid element to cross the ISCO and then return to the main body of the disc, and in fact to some small degree this must happen in a real accretion flow (the ISCO stresses measured in GRMHD simulations are an angular momentum flux sourced from within the ISCO after all). However, the crux of the argument put forward in this paper rests on the assumption that the ISCO acts sufficiently strongly like a one-way gate in the flow that the fluctuation-averaging integrals do not cancel to zero (eq. \ref{fluc_int}).  Given the asymptotic scale separation between radial fluctuations and mean drift, it is only necessary to perturb the precise cancellation in the fluctuation velocity integral to a relatively minor degree for the effects of the boundary to become apparent.  

In purely gravitational dynamics the ISCO is indeed a one-way gate. If a test particle is on a circular orbit at the ISCO, and is perturbed infinitesimally in the radial direction, then it's subsequent radial velocity is given by the solution of the relativistic energy equation $g_{\mu\nu}U^\mu U^\nu = -c^2$, or explicitly \cite{MummeryBalbus22PRL} }
\begin{equation}
    \left[U^r\right]^2 = {2 r_g c^2 \over 3r_I} \left({r_I \over r} - 1\right)^3 . 
\end{equation}
{Here notation is standard, $r_g = GM_\bullet/c^2$. Clearly there are only real solutions of this constraint for $r \leq r_I$ (i.e., inwards perturbations), and the velocity increases rapidly inwards. For a Schwarzschild black hole $U^r$ is already $\sim 0.01 c$ at $r=0.9r_I$, which is a plausible trans-ISCO turbulent perturbation scale in a relatively thin $H/r \sim 0.1$ disc. This radial inflow velocity is already much larger than the typical sound speed in the disc (as will be demonstrated in later sections), and it seems unlikely therefore that the fluid element has a significant probability of returning to the main body of the disc. Of course, if significant non-gravitational forces are present in the flow (for example for the extreme magnetic fields produced in a magnetically arrested disc) this argument may well break down.   While bearing in mind the inherent simplifications employed, we will for the remainder of the paper treat the ISCO as a perfect absorber.  }

\begin{figure}
\centering
\includegraphics[width=0.5\textwidth]{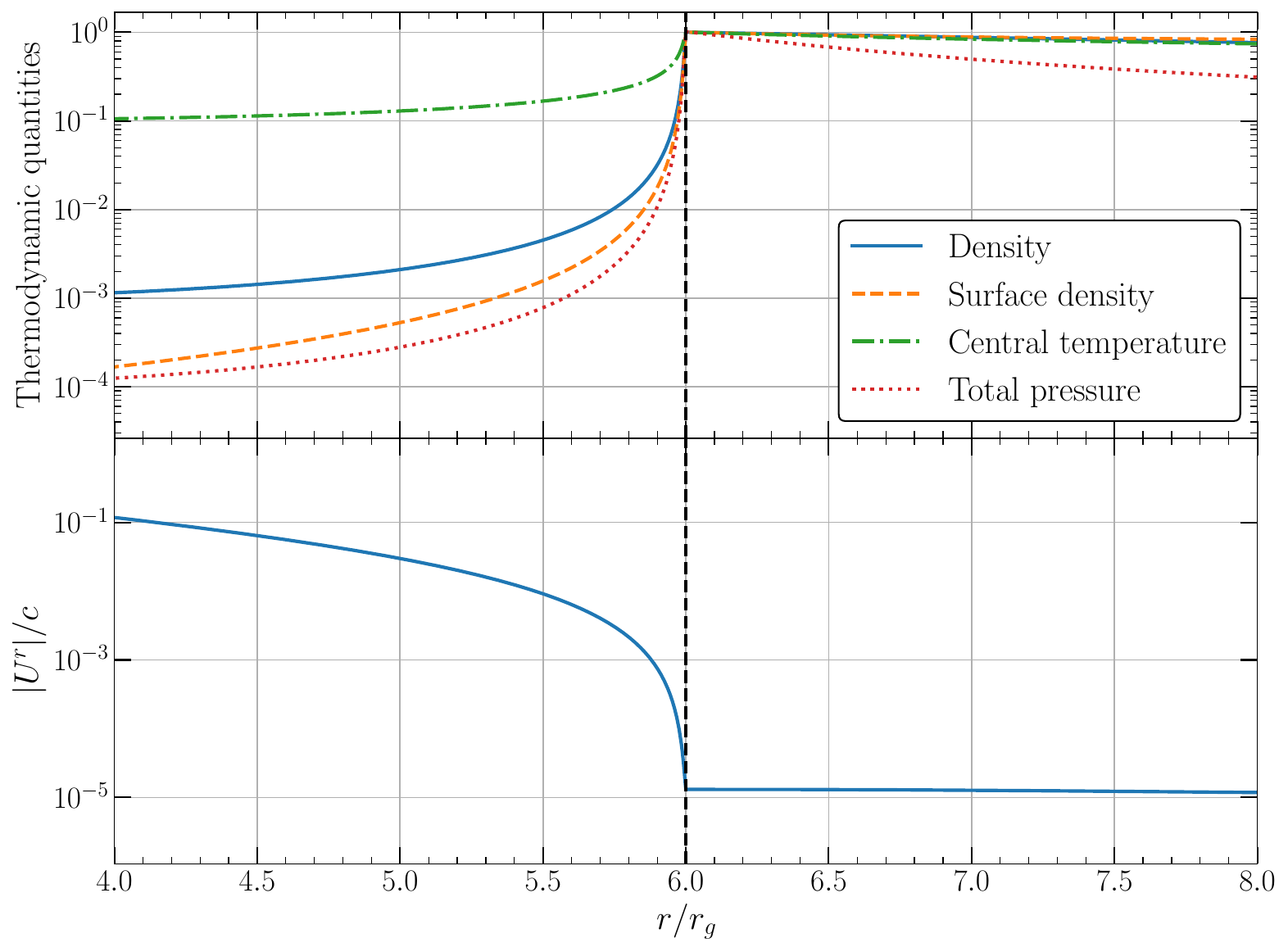}
\caption{ An example of ``cuspy'' behaviour in classical finite ISCO stress models of black hole discs (parameters: $M_\bullet = 10M_\odot, a_\bullet = 0, \dot M = 0.1 \dot M_{\rm edd}, \alpha = 0.1$, $\delta_{\cal J} = 0.1$; see section \ref{implications} for a description of the physical meaning of these free parameters).  In the upper panel we display various normalised (by their ISCO values) disc quantities on both sides of the ISCO (ISCO displayed by vertical black line). The evolution across the ISCO is not smooth in these models, a result of the very low trans-ISCO radial velocity of the disc (lower panel).  We zoom into the inner $2r_g$ either side of the ISCO (black dashed curve, $r_I = 6 r_g$).  }
\label{cusps}
\end{figure}

The trans-ISCO velocity plays a key role in the thermodynamic evolution of disc quantities in the region surrounding the ISCO \citep{MummeryBalbus2023}, and an incorrect value for this velocity can lead to unphysical discontinuous behaviour at the ISCO itself. It turns out that this effect is particularly relevant in the physical regime corresponding to larger ISCO stresses. This is because the mean \cite{NovikovThorne73} flow velocity decreases with ISCO stress $W_I$ \citep[proof in Appendix A of][]{MummeryBalbus2023}
\beq
u_I \propto W_I^{-3/2} .
\eeq
An increased ISCO stress on the other hand causes various disc quantities (for example the central temperature) to increase.  As the acceleration within the ISCO is driven almost entirely by gravity (and therefore is to leading order independent of the local thermodynamics), classical models of finite ISCO stress discs show cuspy behaviour at the ISCO, which is almost certainly unphysical. See Figure \ref{cusps} for an example of this cuspy behaviour.  Indeed, GRMHD simulations generally show a smooth evolution of disc quantities over the ISCO \citep{Shafee08, Noble10, Zhu12, Schnittman16, Liska22}.

The speed of sound, however, increases with a larger ISCO stress 
\begin{equation}
    c_{s, I} \propto W_I^{1/2}, 
\end{equation}
a result of the higher inner disc temperatures. The speed of sound is likely a good measure of the typical turbulent velocity fluctuation scale (in fact this scaling is assumed in classic $\alpha$-models), and therefore while the typical mean drift velocity will decrease, the typical trans-ISCO velocity may well increase as a function of ISCO stress. An increased trans-ISCO velocity in high ISCO stress discs would remove the cuspy behaviour displayed in Fig. \ref{cusps}. The relevant scale of the trans-ISCO velocity is the focus of the analysis in this paper.  

 In effect, the argument we are putting forward here regards the {breakdown} of {classical} ``viscous'' descriptions of turbulent fluids  {when they are within a few turbulent eddy scale lengths of  absorbing boundaries}.  {These classical} descriptions, independent of the magnitude of the {assumed} viscosity, {fundamentally cannot capture behaviour on scales below the eddy length.}  When other physics {(in this case rapid gravitational acceleration)} imprints characteristic length scales on the problem, {this finite length scale of fluctuations will be imprinted in the flow dynamics, which may well be relevant for observational modelling of black hole accretion flows}. 

\section{Random walk models: Theoretical setup and general considerations }\label{rwsettup}
Having identified that the presence of an ISCO (or more generally any absorbing boundary) will act to modify the typical velocity scale with which a fluid element is observed to cross radii close to that boundary, we move to a toy model framework which allows us to probe directly these effects in a controlled manner.  The toy models we turn to are random walks. Random walk systems are well suited to modelling systems with diffusive properties, but also systems with large scale velocity fluctuations, as we now discuss.   

\begin{figure}
\centering 
\includegraphics[width=0.48\textwidth]{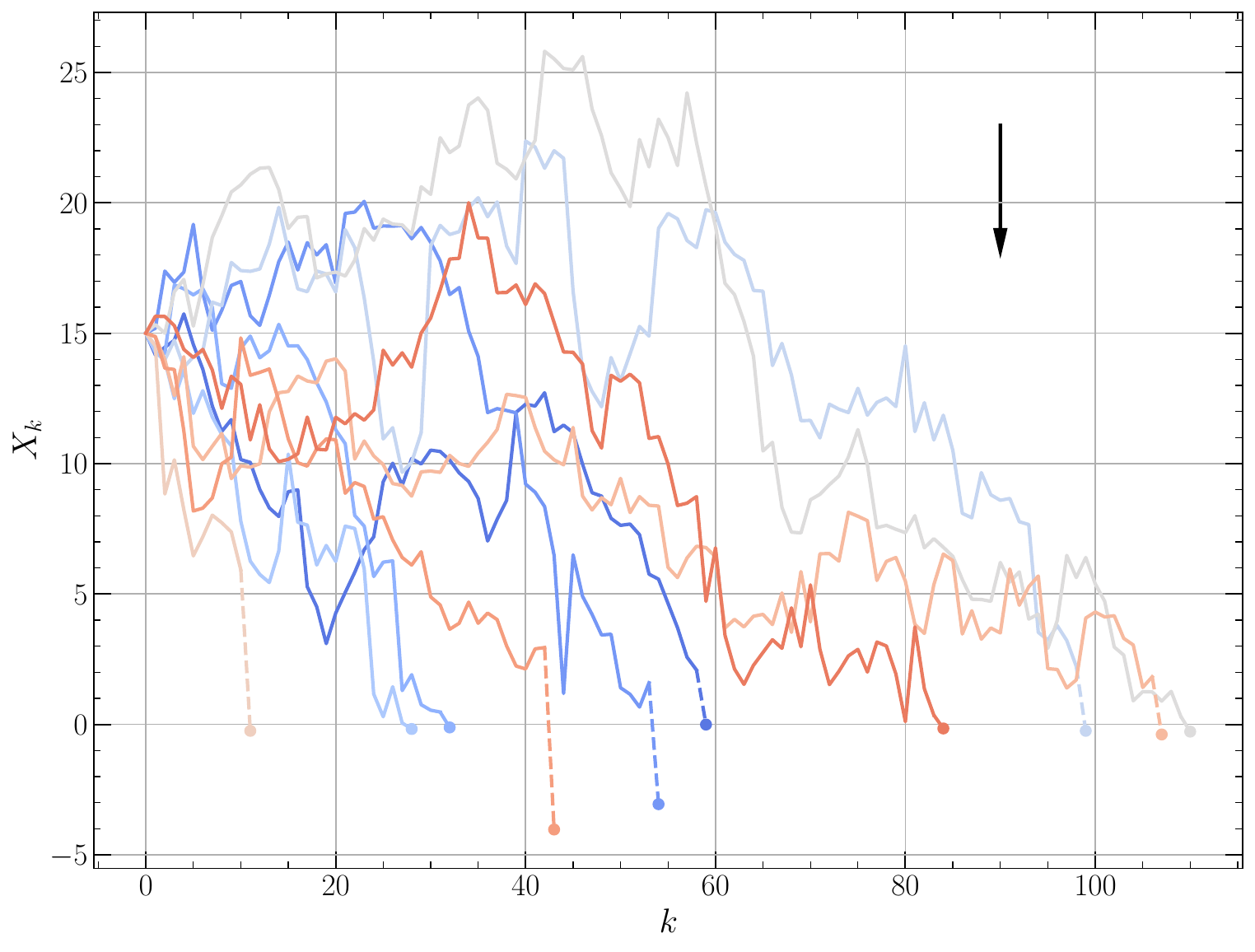}
\includegraphics[width=0.48\textwidth]{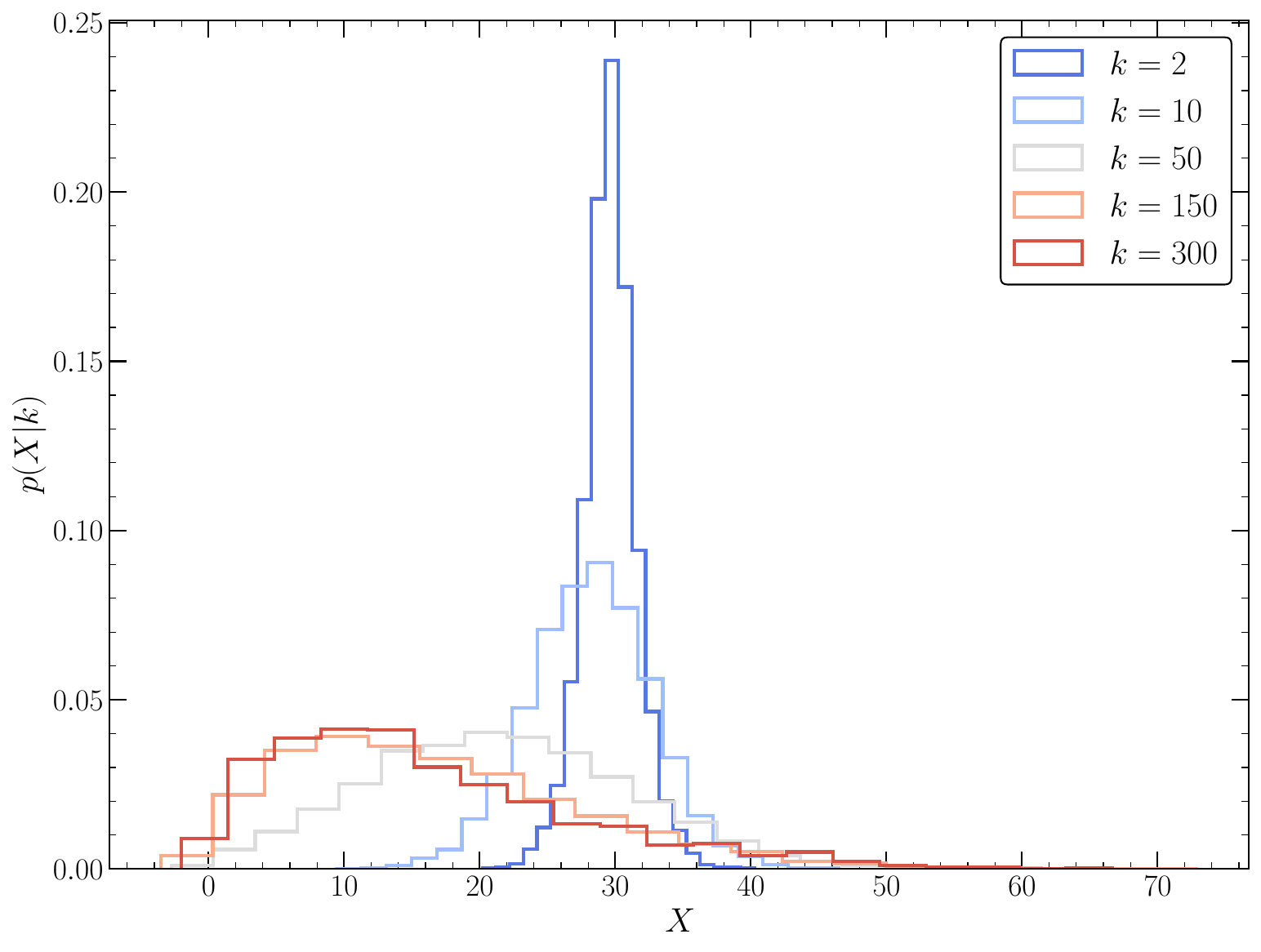}
\caption{Upper panel: an example of individual random walk particle trajectories (solid lines), where we plot the position of the particle $X_k$ after $k$ jumps. The dashed line shows the final jump of the particle, and the dot shows the position of the particle after this last jump. The arrow shows the direction of the imposed drift (the average of the jump distribution, $\epsilon$), ensuring that all particles eventually cross $X = 0$. Clearly random walk models can incorporate the physics of large amplitude velocity fluctuations.  Lower: The evolving (numerical, averaged over $10^4$ trajectories) probability density function of a group of evolving random walk particles at different time-steps $k$. When averaged over large numbers of particles random walk calculations reproduce properties of  diffusive evolution (in the limit of large $k$, with $X_0/\sqrt{k}$ finite).  }
\label{somewalks}
\end{figure}

Consider a one dimensional discrete time random walk. These systems consist of a single particle moving in one dimension, in steps of finite time duration $\Delta t$. The position of a particle currently at position $X_k$ after the next time step is given by 
\beq
X_{k+1} = X_k + v_k \Delta t ,
\eeq
where the velocity of each jump is sampled from some distribution
\beq
v_k \sim f(v). 
\eeq
An example of 10 particle trajectories, which all start at $X_0 = 15$, are shown in the upper panel of Fig. \ref{somewalks}.  For this figure we used the Laplace jump distribution described in later sections (eq. \ref{fv_exp}), which enforces a small drift velocity downwards towards $X = 0$.  For each trajectory the final particle jump is displayed by a dashed line, highlighting the ability of these models to capture large fluctuation dynamics. 

In the lower panel of Figure \ref{somewalks} we highlight the ability of random walk models to capture the diffusive dynamics of systems of many particles. We plot $p(X|k)$ the probability density of finding a hopping particle at a given $X$ coordinate after $k$ steps for an ensemble of $N = 10^4$ hopping particles, which all start at $X = 30$ and evolve with the same Laplacian jump distribution as the upper figure.  This figure highlights the well known result that in the limit $\Delta t \to 0$, random walks asymptote to Brownian motion, and are described by a diffusion equation.  For finite $\Delta t$ the average evolution of many particles still captures many of the properties of diffusive dynamics.  This is an important property of these systems, as we know that the average time-dependent evolution of accreting systems is diffusive (at radii large compared to the ISCO), and therefore random walk models will capture the gross large-radii properties of accreting systems, with the addition of better modelling large amplitude fluctuations in the inner disc.

As we have seen, for a given choice of $f(v)$, and a choice of boundary conditions, a system of many random walk particles can be fully described by some particle density function  $p(x, t)$ which determines the number density of hopping particles in the region $x \rightarrow x + {\rm d}x$ at time $t$. {Keeping in mind the astrophysical setting we are examining, in this notation the coordinate $x$ should be thought of as the distance from the ISCO (i.e., $x=0$ corresponds to the ISCO)}. We wish to answer the question ``what is the velocity distribution of a particle given that it is crossing a boundary at $x = x'$''. We denote this quantity $\widetilde f(v \, | \, x')$. This can be rather generally written (the notation $p(A|B)$ denotes the probability of $A$ given $B$, and we suppress any explicit time dependence) 
\begin{multline}
p(v \, | \, {\rm crossed}\,  x') \equiv \widetilde f(v \, | \, x') = \\ {f(v) \left[ \Theta(v) \int\limits_{x' - v \Delta t}^{x'} p(x) \, {\rm d}x + \Theta(-v)  \int\limits_{x'}^{x' + |v| \Delta t} p(x) \, {\rm d}x\right] \over \int\limits_0^\infty f(v) \int\limits_{x' - v\Delta t}^{x'} p(x) \, {\rm d}x \, {\rm d} v +   \int\limits_{-\infty}^0 f(v) \int\limits_{x'}^{x' + |v|\Delta t} p(x) \, {\rm d}x \, {\rm d} v} ,
\end{multline}
where we split the contributions from positive and negative velocities via the Heaviside theta function $\Theta(z < 0) = 0, \Theta(z > 0) = 1$. In deriving this expression we have made the assumption that each of the particles evolves independently of all other particles. 
In this expression the term on the left of the numerator describes all trajectories moving outwards in the disc ($x' > x$), while the term on the right describes all trajectories moving inwards ($x' < x$). We define the crossing velocity through a surface at radius $x=x'$ as
\begin{multline}
\left\langle v_{\rm cross}(x') \right\rangle \equiv \int\limits_{-\infty}^{+\infty} v \, \widetilde f(v \, | \,   x') \, {\rm d} v \\ =  {\int\limits_{0}^\infty v \, f(v)  \int\limits_{x' - v \Delta t}^{x'} p(x) \, {\rm d}x\, {\rm d}v +  \int\limits_{-\infty}^{0} v \, f(v)  \int\limits_{x'}^{x' + |v| \Delta t} p(x) \, {\rm d}x \, {\rm d}v \over \int\limits_0^\infty f(v) \int\limits_{x' - v\Delta t}^{x'} p(x)\,  {\rm d}x \, {\rm d} v +   \int\limits_{-\infty}^0 f(v) \int\limits_{x'}^{x' + |v|\Delta t} p(x)\,  {\rm d}x \, {\rm d} v} .
\label{vcross}
\end{multline}
Our definition of the crossing velocity is closely related to the ``leapover distance'', which is a well studied quantity in the random walk literature \citep[e.g.,][]{RandomWalk5}. 

 If we have an absorbing boundary at $x' = 0$, for example an ISCO whereafter a fluid element is extremely unlikely to fluctuate back into the stable part of the disc, then $p(x < 0) = 0$\footnote{In an accretion context this statement should be interpreted as the probability of a fluid element fluctuating {\it back} from $x<0$ to $x > 0$ is zero, not that there is zero density there.}, and 
\begin{equation}
\widetilde f(v \, | \,   x'=0)  
=  {f(v)  \Theta(-v)  \int\limits_{0}^{|v| \Delta t} p(x) \, {\rm d}x \over   \int\limits_{-\infty}^0 f(v) \int\limits_{0}^{|v|\Delta t} p(x)\,  {\rm d}x \, {\rm d} v} ,
\end{equation}
and therefore the boundary crossing velocity is
\begin{equation}
\left\langle v_{\rm cross}(x'=0) \right\rangle \equiv \left\langle v_b \right\rangle 
=  {\int\limits_{-\infty}^0 v f(v)   \int\limits_{0}^{|v| \Delta t} p(x) \, {\rm d}x \, {\rm d}v \over   \int\limits_{-\infty}^0 f(v) \int\limits_{0}^{|v|\Delta t} p(x)\,  {\rm d}x \, {\rm d} v} .
\label{vcross_zero}
\end{equation}
It is important to note that only negative velocities contribute to this result (only negative velocities are able to cross the boundary). In addition, higher speeds contribute to a more significant degree than smaller speeds, as they have an increased available crossing distance. We will demonstrate in the following section that explicit models (of $f(v)$) offer insight into the physics we are looking to describe.  

{Finally, it is important to highlight a key assumption we shall be making when solving explicit random walk models. We shall assume that each of the fluid elements in the disc can be treated as undergoing its own random walk, which does not interact with its neighbouring fluid elements. Of course, the dynamics of a real fluid is characterised by complex non-linear interactions, and each fluid element in the disc naturally interacts with its neighbours. The random walk framework is of course an approximation, allowing us to insert ``by hand'' the turbulent fluctuations into the dynamics in a controllable fashion. To leading order, we expect the interactions in a real fluid to make the jump distribution $f$ also a function of the density of particles in the local vicinity and position $f(x, v, p)$. Such a modification would render the random walk problems unsolvable in a closed form, but fortunately we find that our results are insensitive to the precise functional form of $f$, providing confidence in this general approach.  We reiterate that the random walk framework is intended to probe an argument based on physical insight (see section \ref{discussion}), and is not intended to be a quantitative derivation of fluid properties.   }

\section{Explicit models}\label{explicit}
As can be seen in equation \ref{vcross}, the mean crossing velocity at a location $x'$ depends on the probability of finding a particle at all locations $x$ within the disc. The density of particles $p(x, t)$ at position $x$ and discrete time $t = k \Delta t$ evolves according to 
\begin{equation}
p(x,t+\Delta t)=\int\limits_{-\infty}^{\infty} p(x - v \Delta t, t)f(v)\, {\rm d} v .
\end{equation}
Formally the velocity integration limits extend to $\pm \infty$, but constraints can be placed on $f(v)$ to ensure adherence with causality (or relativity, etc.).
In this paper we are interested in steady state models of accreting flows. In particular,  when $t \to \infty$, one finds that the steady-state density $p(x)$ satisfies the integral equation
\begin{equation}
p(x)=\int\limits_{-\infty}^{x/\Delta t} p(x - v\Delta t)f(v)\, {\rm d}v =  \int\limits_{0}^{\infty} p(x')f\left({x-x' \over \Delta t}\right)\, {{\rm d}x' \over \Delta t},
\label{integral_eq}
\end{equation}
where we have defined $x' \equiv x - v\Delta t$, and used  $p(x' < 0) = 0$ to determine the integration limits. This steady state expressions can be interpreted as either the large time behaviour of an initially uniform density of particles $p(x, t=0) = p_0 = {\rm const}$, or the large time behaviour of a system which is continuously fed a stream of particles at a large distance from the inner boundary {(i.e., each new particle starting in the limit $x \to \infty$)}. Equivalently, it can be interpreted as the relative fraction of time a hopping particle starting at large radii $x \to \infty$ spends at each radius on its trajectory to $x = 0$.   This integral equation cannot be solved in general, but we now consider two cases where exact solutions can be found.
\subsection{Simple toy model}
Consider the following model for the distribution of velocities with which a particle can move 
\begin{align}
f(v) &= {1\over 2} \delta(v - v_1) + {1\over 2} \delta(v - v_2), \\ v_1 &= - \epsilon - V, \quad v_2 = -\epsilon + V, \quad V \gg \epsilon. 
\end{align}
This simple model represents a flow with small mean $-\epsilon$ and large fluctuations $V$ which can occur with random (equally probable) direction. This is our simplest first approximation to an accretion flow. Substituting the above jump distribution into the governing integral equation leaves 
\begin{align}
p(x) &= {1\over 2} p(x + \chi)  + {1\over 2} p(x - \beta), \\ \chi &\equiv ( V+ \epsilon )\Delta t, \quad \beta \equiv (V - \epsilon) \Delta t .
\end{align}
This equation has analytical solutions split across two radial regimes, the first where $x < \beta$:
\beq
p_1(x) =  {1\over 2} p_1(x + \chi),
\eeq
with solution $p_1(x) = b_1 \exp(x/D_1)$, and the second where $x \geq \beta$
\beq
p_{2}(x) =  {1\over 2} p_{2}(x + \chi) + {1\over 2} p_{2}(x -\beta ),
\eeq
with solution $p_2(x) =  1-  \exp(-x/D_2)$. The coefficients are given by the solutions of the following equations (determined from the governing equation and simple matching conditions at $x = \beta$) 
\begin{align}
2 &= \exp(-\chi/D_2) + \exp(\beta/D_2) ,\\ 
D_1 &= \chi/\ln(2), \\
b_1 &= \exp(-\beta/D_1) \left[ 1 - \exp(-\beta/D_2)\right] .
\end{align}
Where the upper implicit equation must be solved numerically (Appendix \ref{appB}). 

\begin{figure}
\centering
\includegraphics[width=\linewidth]{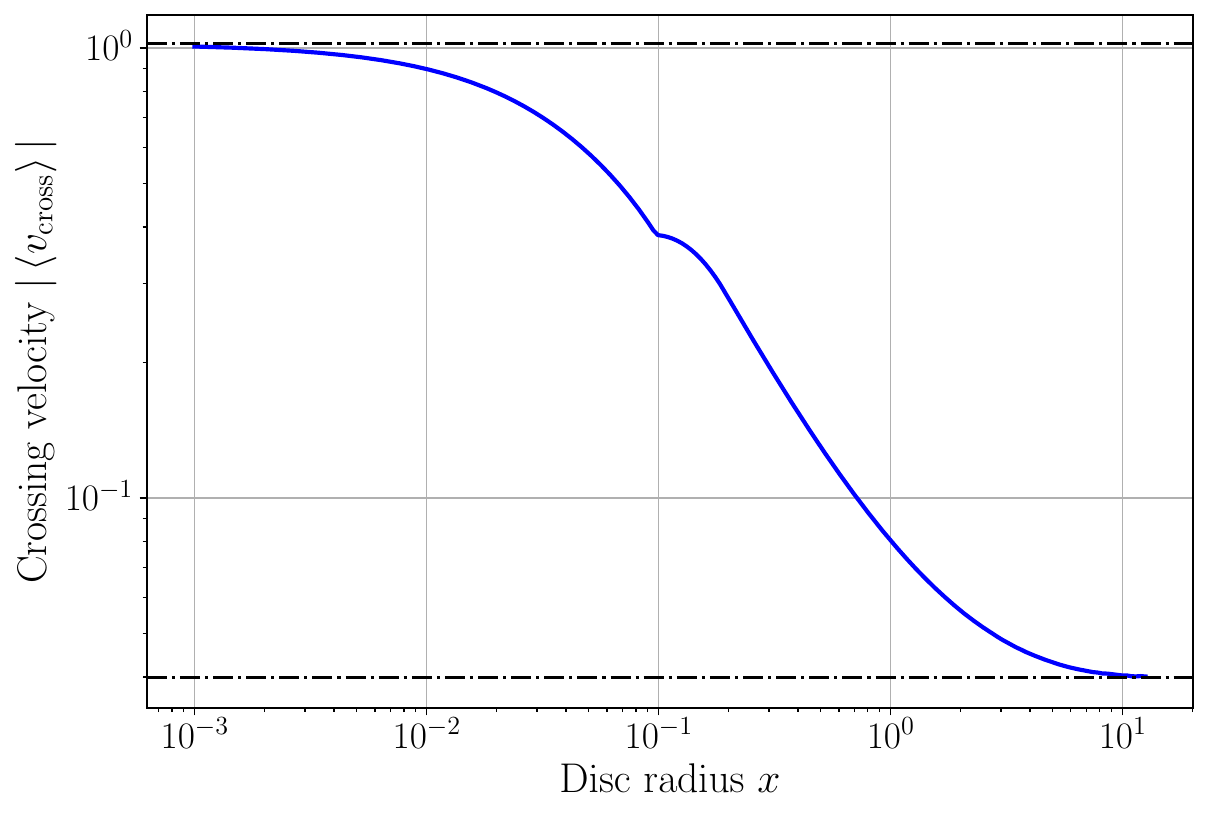}
\caption{The absolute value of the crossing velocity as a function of $x$. The two horizontal black lines are the asymptotic results $|\left\langle v_{\rm cross}(x=0)\right\rangle| = V+\epsilon$ and $|\left\langle v_{\rm cross}(x\gg0)\right\rangle| = 2\epsilon$. For this figure we take $V = 1$, $\epsilon = 0.02$, and time step $\Delta t = 0.1$. The discontinuous gradient of  $\left\langle v_{\rm cross}\right\rangle$ near $x \simeq 0.1$ is due to the discrete nature of the jump distribution $f(v)$. The absorbing boundary is placed at $x = 0$.   }
\label{jump_simple}
\end{figure}

At large distances from the absorbing boundary, the probability of finding a fluctuating particle at a given radius will tend to unity $p(x\gg D_2) \to 1$, therefore  
\begin{multline}
\left\langle v_{\rm cross}(x'\gg 0) \right\rangle \simeq  {\int_{0}^\infty v \, f(v) v \Delta t  \, {\rm d}v +  \int_{-\infty}^{0} v \, f(v)  |v| \Delta t\, {\rm d}v \over \int_0^\infty f(v) v\Delta t \, {\rm d} v +   \int_{-\infty}^0 f(v) |v|\Delta t \, {\rm d} v} \\= {v_2^2 - v_1^2 \over v_2 - v_1 } = v_2 + v_1 =  - 2\epsilon.
\end{multline}
whereas at $x' = 0$, we have
\begin{multline}
\left\langle v_{\rm cross}(x'=0) \right\rangle=  {\int_{-\infty}^0 v f(v)   \int_{0}^{|v| \Delta t} p(x) \, {\rm d}x \, {\rm d}v \over   \int_{-\infty}^0 f(v) \int_{0}^{|v|\Delta t} p(x) {\rm d}x \, {\rm d} v} \\ = {{1\over 2} v_1 \int_0^{|v_1|\Delta t} p(x) \, {\rm d}x \over {1\over 2} \int_0^{|v_1| \Delta t} p(x) \, {\rm d}x } = v_1 = -\epsilon -V . 
\end{multline}
Both of these results are readily interpretable.  Firstly, at large radii, the typical surface crossing velocity is twice the mean drift velocity of the particle. The fact that the crossing velocity is order the mean drift velocity is an entirely expected result. The factor two simply results from the smaller distance on the positive jump side from which particles can cross a given boundary (as the mean and fluctuating velocities are working against each other) compared to the negative jump direction (where they work together). On the other hand, the inner boundary crossing velocity is given simply, and intuitively,  by the velocity of the negative jump. In Fig. \ref{jump_simple} we display the crossing velocity as a function of location $x$, showing the clear increase in $\left\langle v_{\rm cross} \right\rangle$ as the absorbing boundary is approached.


\subsection{ An exponential random walk  }
The delta-function jump distribution considered previously naturally represents an unphysical simplification.  In this sub-section we demonstrate that the key results are unchanged by considering a more complex jump distribution. 

Consider the case of the exponential random walk, with jump velocity  distribution given by the Laplace distribution \citep[e.g.,][]{RandomWalk1, RandomWalk2}
\begin{equation}
f(v)=\frac{1}{2V}\exp\left(-{|v+\epsilon|\over V}\right)\,,
\label{fv_exp}
\end{equation}
where $\epsilon>0$ is the magnitude of the drift (in the direction of lowering $x$) $,\sqrt{2}V$ is the standard deviation of the noise and $|z|$ represents the absolute value of a variable $z$.  This specific jump distribution $f(v)$ has the following property
\begin{equation}
{{\rm d} ^2 \over {\rm d}v^2}f(v)=\frac{1}{V^2}f(v)-\frac{1}{V^2}\delta(v+\epsilon)\,,
\label{trick}
\end{equation}
which will be useful to compute the steady state properties.

\begin{figure}
\centering
\includegraphics[width=\linewidth]{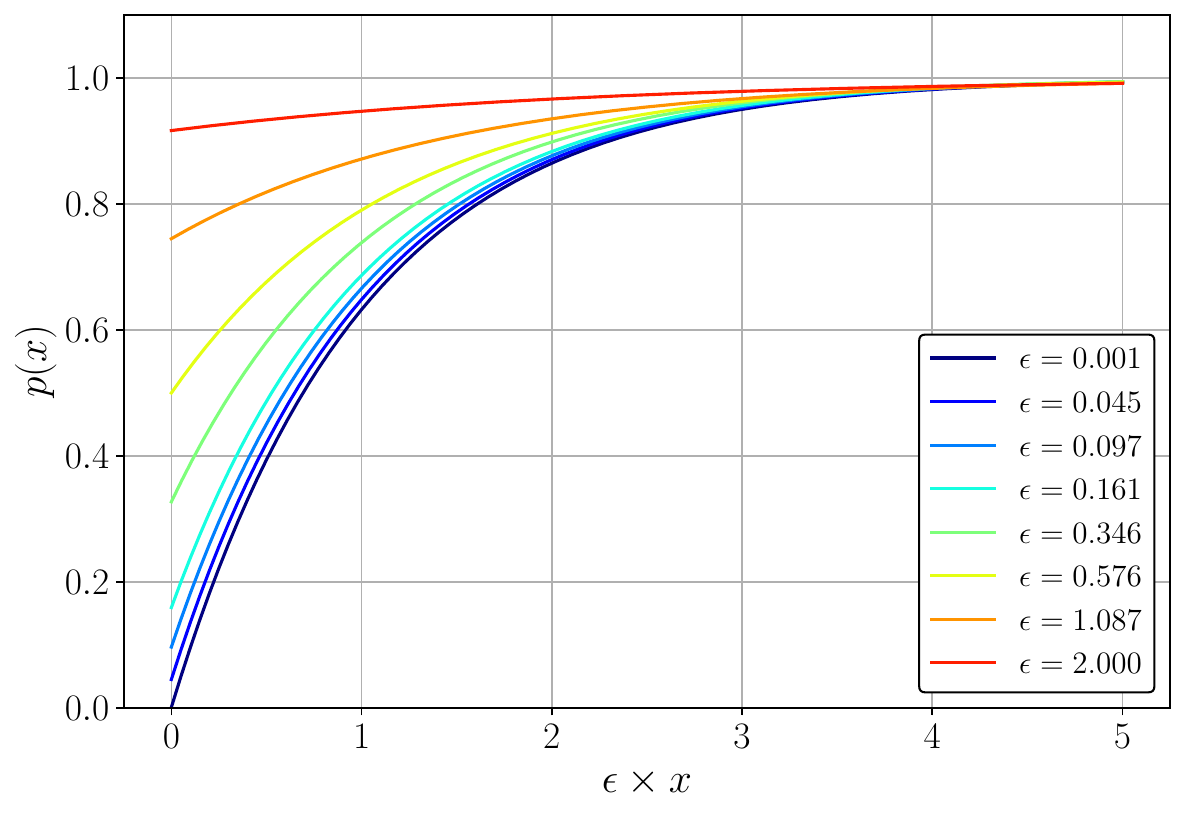}
\caption{The probability density function $p(x)$ as a function of $\epsilon x$ for different values of the drift velocity $\epsilon$. For this Figure the fluctuation velocity scale was normalised to $V = 1$, and the time step was $\Delta t = 0.1$. The absorbing boundary is placed at $x = 0$.   }
\label{p_exp}
\end{figure}

We can make progress by taking two derivatives with respect to $x$ on both sides of Eq.~\eqref{integral_eq}, yielding
\begin{equation}
{{\rm d}^2 p\over {\rm d}x^2} = {1\over \Delta t} \int\limits_{0}^{\infty} p(x') {{\rm d}^2  \over {\rm d}x^2}\left[f\left({x-x' \over \Delta t}\right) \right] \,{\rm d}x'.
\label{integral_eq_2}
\end{equation}
Using the property in Eq.~\eqref{trick}, we find
\begin{equation}
{{\rm d}^2 p\over {\rm d}x^2}=\frac{1}{(V\Delta t)^2}\left[p(x)-p(x+\epsilon\Delta t)\right]\,.
\label{diff_eq}
\end{equation}
In other words, starting from an integral equation, we have derived a nonlocal differential equation for $p(x)$.  
It will again be of use to define the two length scales $\chi \equiv \epsilon \Delta t$ and $\beta \equiv V \Delta t$. We then make the ansatz \citep{RandomWalk3}
\beq
p(x) = 1 + b \exp(-x/D) ,
\eeq
which upon substitution leads to the following implicit equation for the length scale $D$
\beq
\left({\beta \over D}\right)^2 = 1 - \exp\left(-{\chi \over D}\right),
\eeq
which has a unique solution  $D > 0$ for every $\chi, \beta > 0$ (Appendix \ref{appB}).  Note that in the limit of small drift velocities $\epsilon \to 0$ (the relevant limit for accretion), we have 
\beq\label{inner_scale}
D \to \delta =  {\beta ^2 \over \chi } = {V^2 \Delta t \over \epsilon} , 
\eeq
which can be thought of as  the length scale at which the flow starts to ``learn'' about the boundary.
Inserting this expression for $p(x)$ back into the integral equation \eqref{integral_eq}, we find 
\begin{equation}
1 + b e^{-x/D} = {1 \over \Delta t} \int\limits_{0}^{\infty} \left(1+be^{- x'/D}\right)f\left({x-x' \over \Delta t}\right) \, {\rm d}x',
\label{integral_eq_3}
\end{equation}
which gives a condition to determine $b$. This determination is most easily performed at $x = 0$. Explicitly, 
\begin{multline}
b \left(1 - {1 \over 2 \beta } \int\limits_{0}^{\infty} \exp\left(- {|\chi - x'| \over \beta} - {x' \over D}\right) \, {\rm d}x'\right) \\ = -1 + {1 \over 2 \beta } \int\limits_{0}^{\infty} \exp\left(- {|\chi - x'| \over \beta} \right) \, {\rm d}x' .
\end{multline}
Each integral is now elementary (see Appendix \ref{appA}), and we have found an exact solution for the steady state probability density function of a discrete time random walk. We plot the solution $p(x)$ for different values of $\epsilon$ in Fig. \ref{p_exp} {(including both the astrophysically relevant $\epsilon \ll V$ limit, but also larger $\epsilon$)}. The smaller the drift velocity the earlier the flow ``learns'' about the absorbing boundary. {This can be seen in  Fig. \ref{p_exp}, where each solution differs from the constant $p(x) \simeq 1$ at $\epsilon x \simeq 3$, meaning that the length scale at which the flow begins to ``learn'' about the boundary scales as $\delta \sim 1/\epsilon$, as predicted from eq. (\ref{inner_scale}). }

\begin{figure}
\centering
\includegraphics[width=\linewidth]{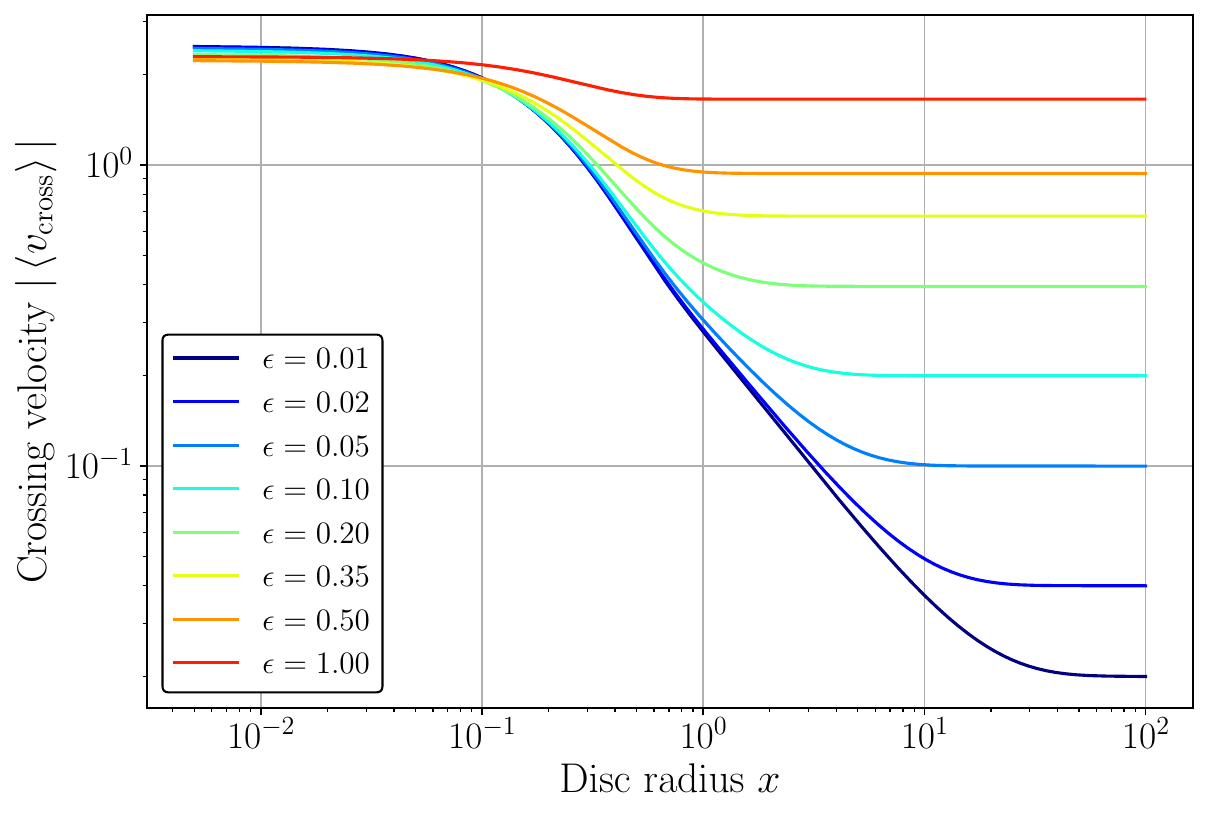}
\caption{The absolute value of the crossing velocity $\left\langle v_{\rm cross}\right\rangle$ as a function of disc radius $x$ for different values of the drift velocity $\epsilon$. For this Figure the fluctuation velocity scale was normalised to $V = 1$, and the time step was $\Delta t = 0.1$.  The absorbing boundary is placed at $x = 0$.  }
\label{cross_vel_exp}
\end{figure}

With the probability density $p(x)$ determined, the crossing velocity may be found straightforwardly from equation (\ref{vcross}).  We display the crossing velocity as a function of $x$ for the exponential jump distribution in Fig. (\ref{cross_vel_exp}), for a number of different drift velocities $\epsilon$ {(including both the astrophysically relevant $\epsilon \ll V$ limit, but also larger $\epsilon$)}. As expected, far from the inner boundary the crossing velocity is of order $\epsilon$, and independent of the typical velocity fluctuation scale $V$.  However, as the inner boundary is approached, the crossing velocity increases substantially and becomes effectively independent of the drift velocity $\epsilon$ at the location of the boundary edge.   

\begin{figure}
\centering
\includegraphics[width=\linewidth]{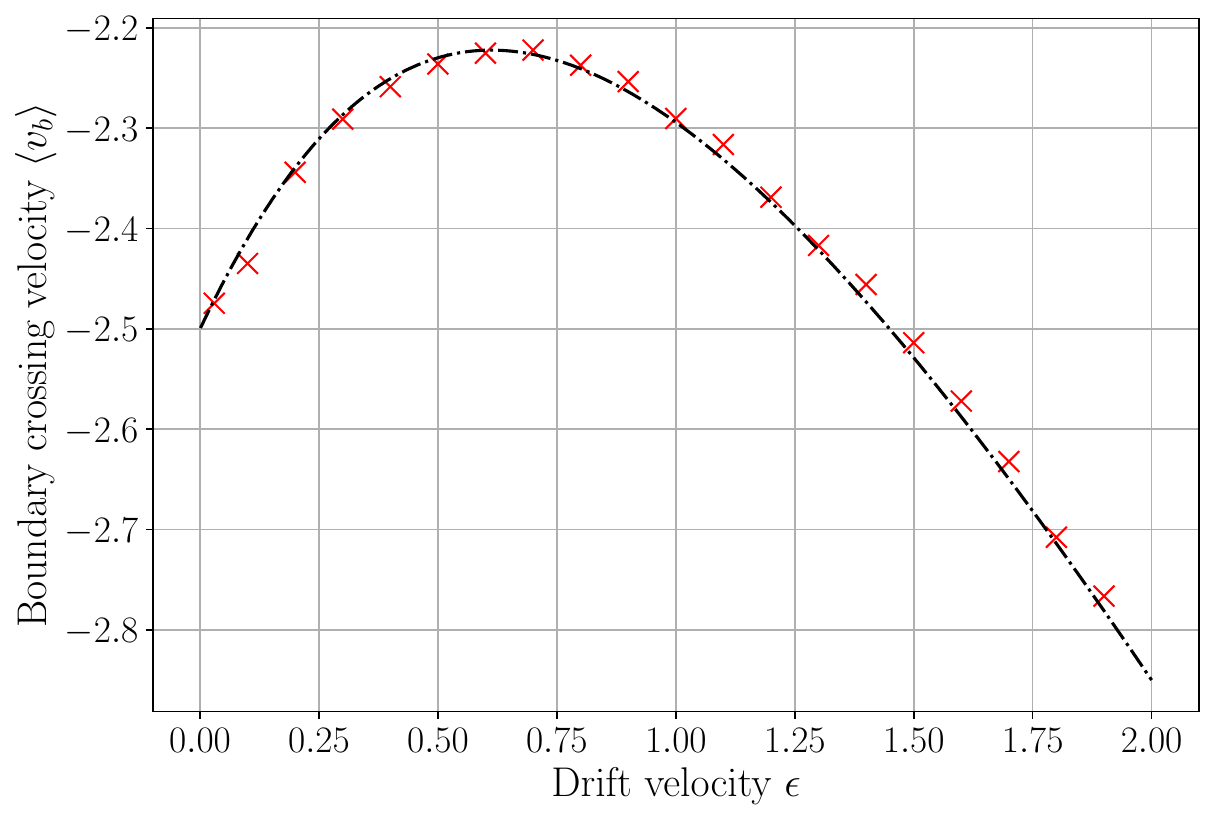}
\caption{The inner boundary crossing velocity, $\left\langle v_b\right\rangle$, as a function of the mean drift velocity $\epsilon$. For this Figure the fluctuation velocity scale was normalised to $V = 1$. The black dot-dashed curve is the analytical result derived in this paper, whereas the red crosses are numerical averages of $N = 10^5$ individual particle trajectories. In the small drift velocity regime relevant for accretion the boundary crossing velocity is always  $\left\langle v_b\right\rangle \gg \epsilon$.   }
\label{cross_vel_inner}
\end{figure}

The crossing velocity across the absorbing boundary can be computed exactly, and is given by the following expression  
\begin{multline}
\left\langle v_b \right\rangle 
= {\int\limits_{-\infty}^0 v f(v)   \int\limits_{0}^{|v| \Delta t} 1 + b e^{-x/D} \, {\rm d}x \, {\rm d}v \over   \int\limits_{-\infty}^0 f(v) \int\limits_{0}^{|v|\Delta t} 1 + b e^{-x/D}\,  {\rm d}x \, {\rm d} v} \\
= {\int\limits_{-\infty}^0 v (bD -v\Delta t - bD \exp\left({v\Delta t \over D}\right)) \exp\left(-{|v+\epsilon|\over V}\right)   \, {\rm d}v \over   \int\limits_{-\infty}^0  (bD -v\Delta t - bD \exp\left({v\Delta t \over D}\right))  \exp\left(-{|v+\epsilon|\over V}\right) \, {\rm d} v} ,
\label{vcross_zero_exact}
\end{multline}
which is also an elementary integral (presented in full in Appendix \ref{appA}). The relevant result in the accretion context is the $\epsilon \to 0$ limit, where we find 
\beq
\left\langle v_b \right\rangle = -{5\over 2 } V + \epsilon - {\cal O}(\epsilon^2) .
\eeq
We find that the inner boundary crossing velocity is always significantly larger than $\epsilon$ when $\epsilon \ll V$. Figure \ref{cross_vel_inner} shows the inner boundary crossing velocity as a function of drift velocity\footnote{The turning point observed in the boundary crossing velocity is an example of a phenomena known as ``negative differential mobility'', which is observed in some, but not all, random walk systems \citep{RandomWalk4}.}, including results from numerical simulations.  {The potentially counter intuitive result that the boundary crossing velocity is more than twice the typical fluctuation scale can be understood by noting that particles travelling towards the boundary with higher speeds contribute to a more significant degree than those crossing with smaller speeds, as they have an increased available crossing distance and are therefore preferentially selected.  }

\subsection{Crossing velocity far away from the boundary}
It is important to verify that the crossing velocity as defined in this work has the ``proper'' behaviour at large radii, i.e., that it tends to a constant value corresponding to the mean drift of the flow, independent of the properties of the (unknown) jump distribution, $f(v)$. 

Far away from the absorbing boundary the stationary density $p(x)$ is approximately constant (as there is no length scale in the problem). As a consequence, from Eq.~\eqref{vcross} we find
\beq
\left\langle v_{\rm cross}(x\to \infty) \right\rangle \approx \frac{\int\limits_{-\infty}^{\infty}v |v| f(v)\, {\rm d}v}{\int\limits_{-\infty}^{\infty} |v| f(v) \, {\rm d}v}\,.
\eeq
We perform the change of variable $v\to v'=v+\epsilon$, where $-\epsilon$ is  the mean of the jump distribution $f(v)$, here assumed small compared to the fluctuation scale. This yields 
\begin{multline}
\left\langle v_{\rm cross}(x\to \infty) \right\rangle \approx \frac{\int\limits_{-\infty}^{\infty}(v'-\epsilon) |v'-\epsilon| f(v'-\epsilon)\,{\rm d} v'}{\int\limits_{-\infty}^{\infty} |v'-\epsilon| f(v'-\epsilon)\,{\rm d}v'}\\ =\frac{\langle (v'-\epsilon)|v'-\epsilon|\rangle_{v'}}{\langle |v'-\epsilon|\rangle_{v'}}\,,
\end{multline}
where the notation $\langle . \rangle_{v'}$ denotes an average over $v'$. Note that $\langle v'\rangle_{v'}=0$ by construction. We also assume that the distribution of $v$ is symmetric around $v=-\epsilon$, i.e., that $f(v'-\epsilon)=f(-v'-\epsilon)$. In the limit where the variance of $f(v)$ is much larger than $\epsilon$, the contribution to these integrals will come from velocities $|v'| \gg \epsilon$, and therefore we may expand $|v'-\epsilon|\approx |v'|-\operatorname{sgn}(v')\epsilon$. This leaves 
\begin{multline}
\left\langle v_{\rm cross}(x\to \infty) \right\rangle \approx \frac{\langle (v'-\epsilon)(|v'|-\epsilon \operatorname{sgn}(v'))\rangle_{v'}}{\langle |v'|-\epsilon \operatorname{sgn}(v')\rangle_{v'}} \\ \approx \frac{-2\epsilon \langle |v'| \rangle_{v'}}{\langle |v'| \rangle_{v'}}=-2\epsilon\,,
\end{multline}
where we have used the fact that $\langle |v'|v'\rangle_{v'}=0$ and $\langle \operatorname{sgn}(v')\rangle_{v'}=0$, as a consequence of the symmetry of the distribution of $v'$. Note that for distributions with $\langle v'\rangle_{v'}=0$ but $f(v'-\epsilon)\neq f(-v'-\epsilon)$ the result above would not be valid in general.

\section{Implications of these results in an accretion context}\label{implications}
Both the toy random walk calculations, and the general insight of section \ref{discussion}, suggest that the typical velocity with which a fluid element crosses the ISCO is much better approximated by the turbulent velocity scale than the mean drift velocity of the flow.   In this section we highlight the effects this insight has on the thermodynamic disc quantities on either side of the ISCO.  

\subsection{Typical scales of relevant parameters }
In this section we estimate the relevant scales of the fluctuation and drift velocities in a standard accretion flow.  We perform this analysis in the Newtonian limit, as the idea is only to understand the scales involved, not perform a full rigorous analysis.  We follow the notation of section \ref{explicit}, where we denote by $\epsilon$ the (small) radial drift velocity of the flow, and $V$ the typical turbulent velocity fluctuation scale.  

Using standard $\alpha$-type scaling arguments, we note that the typical velocity fluctuation scale is assumed to be 
\beq
V \sim \alpha^{1/2} \, c_s \sim \alpha^{1/2} \left({H \over R} \right) v_\phi.  
\eeq
This follows from the definition of the usual \cite{SS73} alpha prescription 
\beq
W^{r\phi} = \left\langle \delta v^r \delta v^\phi \right\rangle \sim \alpha c_s^2 , 
\eeq
and by assuming $\delta v^r$ and $\delta v^\phi$ have similar magnitudes. 
The typical drift velocity is \citep[e.g.,][]{Pringle81} 
\beq
 \epsilon \sim \alpha \left({H \over R}\right)^2 v_\phi , 
 \eeq
 and the fluctuation timescale is of order the orbital timescale (i.e., turbulent fluctuations are excited over the shortest timescale in the problem)
 \beq
  \Delta t \sim t_{\rm orb} \sim {R \over v_\phi} . 
\eeq
In these expressions we have used the approximate (Newtonian) solution of hydrostatic equilibrium to relate the sound speed to the orbital speed 
\beq
c_s = v_\phi \left({H \over R}\right) .
\eeq
In the accretion context we are therefore well into the $\epsilon/V \to 0$ limit:
\beq
\epsilon/V \sim \alpha^{1/2} \left({H \over R}\right) \sim 10^{-2} ,
\eeq
for typical (thin) disc parameters. 

The toy exponential jump distribution model of section \ref{explicit} demonstrates that there is a radial scale at which the crossing velocity begins to deviate from the drift velocity, or in effect there is a radial scale at which the flow starts to ``learn'' of the absorbing boundary.  Using the scaling highlighted by eq. (\ref{inner_scale}), we find 
\beq
\delta = {V^2 \Delta t \over \epsilon} \sim R ,
\eeq
i.e., the flow starts to learn of the ISCO roughly $\sim$ one ISCO radius away from the ISCO. Within an innermost layer of size (see fig. \ref{cross_vel_exp}) 
\beq
\delta_{\rm in} \sim V\Delta t \sim \alpha^{1/2} \, H, 
\eeq 
the crossing velocity is given by its $x\to 0$ asymptotic value of $\sim \alpha^{1/2} c_s$. 

\subsection{Thermodynamic solutions near to the ISCO }

\subsubsection{Energy conservation }
We have argued in this paper that the classical calculation of the radial velocity of an accretion flow must be modified near to the ISCO radius, as the turbulent velocity fluctuations, assumed to vanish in the main body of the disc on average, develop a non-zero directional bias as a result of the absorbing boundary in the flow. To characterise the thermodynamic properties of the flow one begins by solving the constraints of energy conservation. It is interesting to highlight how a non-zero directional bias in $\delta U^r$, provided it remains smaller than the orbital velocity scales, does not modify the dominant energy balance equation, and therefore the classical \citep{NovikovThorne73, PageThorne74} temperature profile of the disc is unchanged. 

The energy balance in the main body of the disc can be determined from the conservation of the stress-energy tensor of the disc  $T^{\mu\nu}$. We present a full relativistic calculation of this energy balance in Appendix \ref{appC}, while here simply quoting the key result, namely: 
\begin{multline}\label{Ch2econ3}
\rho \left(U^r + \delta U^r\right) \left[ U^0 \partial_r U_0 + U^\phi \partial_r U_\phi \right] +  {U^0 \over \sqrt{g}} \partial_\mu \left(\sqrt{g} \rho W^\mu_0 \right) \\ + {U^\phi \over \sqrt{g}} \partial_\mu \left(\sqrt{g} \rho W^\mu_\phi \right)   = -(U^0U_0 + U^\phi U_\phi )  \partial_z q^z ,
\end{multline}
where $\rho$ is the density of the disc, $W^{\mu\nu}$ is the turbulent stress tensor, and $q^z$ is the heat radiated out of the disc surfaces by photons. 

We see that there is a term proportional to $U^r + \delta U^r$ in this energy balance equation, which describes the effects of advection,  and so at first it may appear that the work in this paper modifies the temperature profiles of standard theory. However, the prefactor of the advection term is itself of order the radial velocity scale 
\beq
U^0 \partial_r U_0 + U^\phi \partial_r U_\phi \sim {\cal O}\left(U^r + \delta U^r \right) ,
\eeq
as it vanishes identically in the Kerr midplane for circular orbits \citep[see Appendix 3 of][for a formal proof]{MumBalb19a}.  

As such, modifications to $U^r + \delta U^r$ which remain sub-orbital $U^r + \delta U^r \ll r U^\phi$ do not substantially modify the energy balance constraint, as the advection term enters at order ${\cal O}\left(U^r/r U^\phi\right)$. The above identity will cease to be true for extreme values of $U^r$, when the energy balance becomes advection dominated \citep[and a ``slim disc'' regime is entered][]{Abram88}. However, we shall show that the typical $\alpha^{1/2} c_s$ values of thin discs at moderate accretion rates are of order $\sim 10^{-3} c$, while the rotational velocities at the ISCO are of order $rU^\phi \sim c$, meaning this reasoning is robust. Within the ISCO, of course, this argument completely breaks down and a new formalism must be employed \citep{MummeryBalbus2023}.

The final steps of the derivation of the radiative temperature of these disc solutions are therefore unchanged from the classic \citep[][]{NovikovThorne73, PageThorne74} calculation, and we do not repeat them here. The final result, in the steady state, is that the radiative temperature $T_R$ depends on the remaining parameters through 
\beq
\sigma T_R^4 = {3GM_\bullet \dot M \over 8\pi r^3} \big[{\cal R}(\eta) - (1- \delta_{\cal J}){\cal R}(\eta_I)  \big]  \left[ 1 -{ 3 \over  \eta^{2} } + { 2 a_\bullet \over \eta^{3}} \right]^{-1} ,
\eeq
where $M_\bullet, a_\bullet, \dot M$ are the black hole mass, dimensionless spin parameter and disc accretion rate respectively, $\eta$ is the square root of the radius normalised to the gravitational radius $\eta \equiv \sqrt{r c^2 / GM_\bullet}$, and ${\cal R}(\eta)$ is the relativistic correction function due to \cite{PageThorne74}
\begin{equation}
    {\cal R}(\eta) = 1 - {3a_\bullet \over 2 \eta} \ln(\eta) + {1\over \eta} \sum_{\lambda = 0}^{2} k_\lambda \ln\left|\eta - \eta_\lambda\right| ,
\end{equation}
where 
\beq
\eta_\lambda = 2 \cos\left[ {1\over 3} \cos^{-1}(-a_\bullet) - {2\pi\lambda\over3}\right] 
\eeq
and 
\beq
 k_\lambda \equiv {2 \eta_\lambda - a_\bullet(1 + \eta_\lambda^2)  \over 2(1 - \eta_\lambda^2)} .
 \eeq
The parameter $\delta_{\cal J}$ expresses the inner boundary condition of the ISCO stress, where $\delta_{\cal J} = 0$ represents a vanishing ISCO stress.  The parameter $\delta_{\cal J}$  corresponds physically to the fraction of its ISCO angular momentum an accreting fluid element is able to pass back to the main body of the disc over its plunge. 

\subsubsection{Solving for the thermodynamic profiles }
As we have just demonstrated, upon specifying the free parameters of the disc theory ($\dot M, M_\bullet, a_\bullet, \alpha, \delta_{\cal J}$) we have a predetermined  radiative temperature profile $T_R(r)$. We now derive the full solutions of the thermodynamic disc properties with a modified radial velocity profile.

\begin{figure*}
\includegraphics[width=\textwidth]{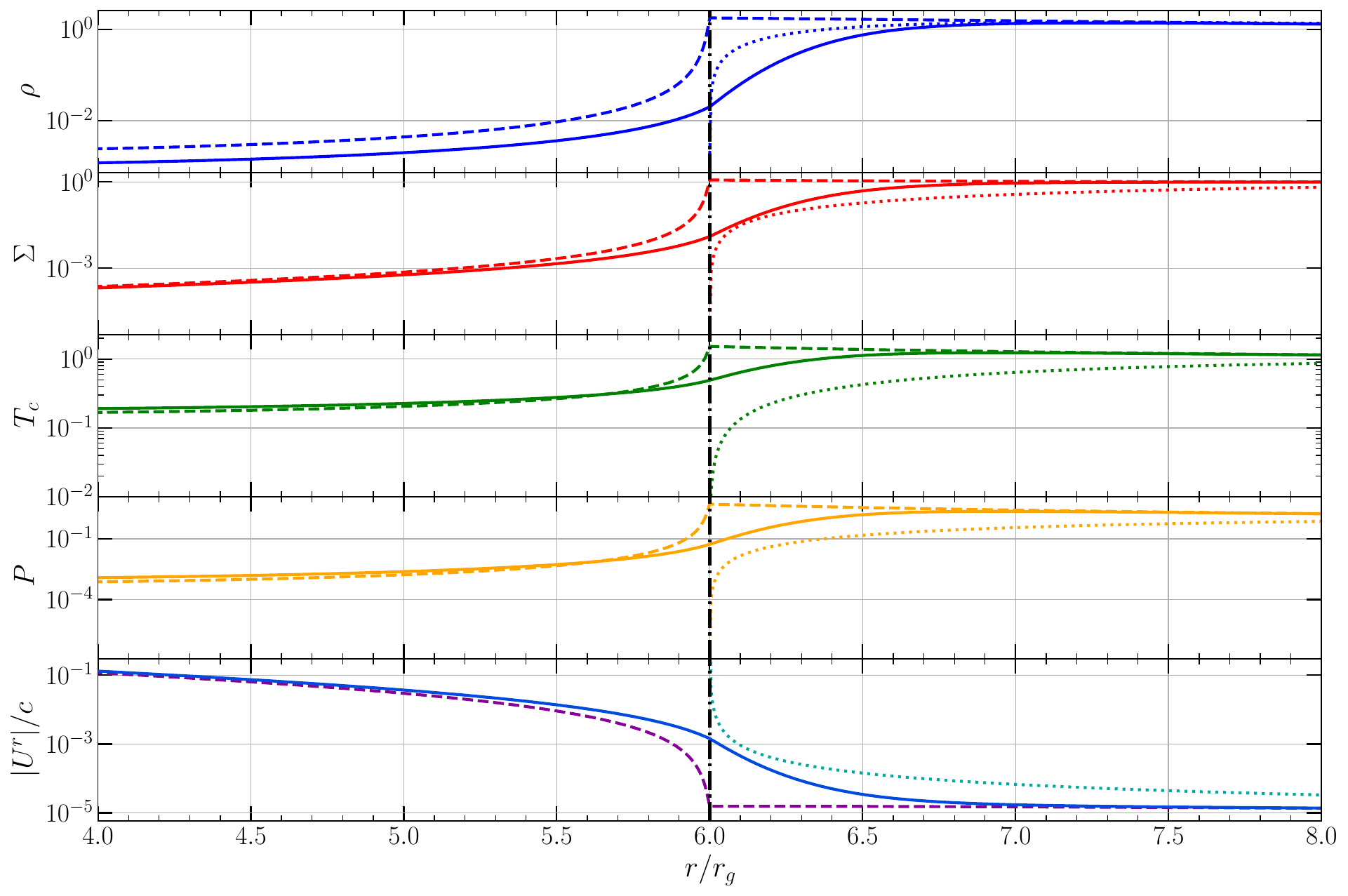}
\caption{Various disc thermodynamic quantities, normalised by their values at $10r_g$, for differing models of the inner disc flow. We display with dashed curves the classical finite ISCO stress solutions, which display pronounced cusps at the ISCO, and by dotted curves the corresponding vanishing ISCO stress solution, which show an unphysical radial velocity dispersion $U^r \to \infty$ at the ISCO.  Finally, by solid curves we display the new solutions derived in this work, with a trans-ISCO velocity set by $\alpha^{1/2} c_s$, we see that the cuspy nature of the finite ISCO stress solutions has been removed, and the transition across the ISCO is significantly smoother. In the lowest panel we show the radial velocities of the three models, normalised by the speed of light.  The ISCO radius is denoted by a vertical black dot-dashed line.    }
\label{Therm}
\end{figure*}

From mass conservation the radial velocity is related to the surface density through 
\beq
\dot M = 2\pi r \Sigma U^r . 
\eeq
Vertical hydrostatic equilibrium gives the scale height of the disc \citep[][]{Abramowicz97}
\beq
H =  \sqrt{P r^4 \over \rho (U_\phi^2 + a^2 c^2 (1 - U_0^2))}  ,
\eeq
which relates the discs surface density $\Sigma$ to the disc density $\rho$ 
\beq
\rho \equiv {\Sigma \over H} .
\eeq
The pressure of the disc is given by the sum of the gas and radiation pressures 
\beq
P = P_g + P_r = {\rho k T_c \over \mu m_p} + {4\sigma T_c^4 \over 3 c} . 
\eeq
Finally, the central and radiative temperatures are related through the optical depth 
\beq
T_c^4 = {3\over 8} \kappa \Sigma T_R^4,
\eeq
and we shall assume that electron scattering opacity dominates within the flow $\kappa \simeq \kappa_{\rm es}$. 

All of the above expressions are completely standard.  The new addition, using the insight gained from the random walk calculations,   is that the trans-ISCO velocity of the flow will be 
\beq
U^r(r_I) = -\alpha^{1/2} c_{s, I} , 
\eeq
where $c_{s, I}$ is the ISCO speed of sound. It turns out that this is sufficient to close the full set of disc equations at the ISCO. Remembering that $T_{R}(r_I) \equiv T_{R, I}$ is known, the above equations can be manipulated into  an algebraic equation  for $c_{s, I}$ in terms of $T_{R, I}, \dot M, M, a$ and $\alpha$.  Explicitly, the ISCO speed of sound satisfies (see Appendix \ref{appD} for a derivation) 
\begin{equation}
 {\sigma \kappa T_{R, I}^4  \over 2 c} {1\over  c_{s, I}} \sqrt{r_I^3 \over 2 G M_\bullet}+ \left({3\kappa\dot M \over 16  \pi r_I \alpha^{1/2}}\right)^{1/4} {k T_{R, I}\over \mu m_p}{1 \over c^{9/4}_{s, I}} = 1 ,
\end{equation}
which is trivial to solve numerically. With this boundary condition determined, the remaining thermodynamic disc profiles can be computed once a specification of the disc's radial velocity is given. 

\begin{figure*}
\centering
\includegraphics[width=0.45\textwidth]{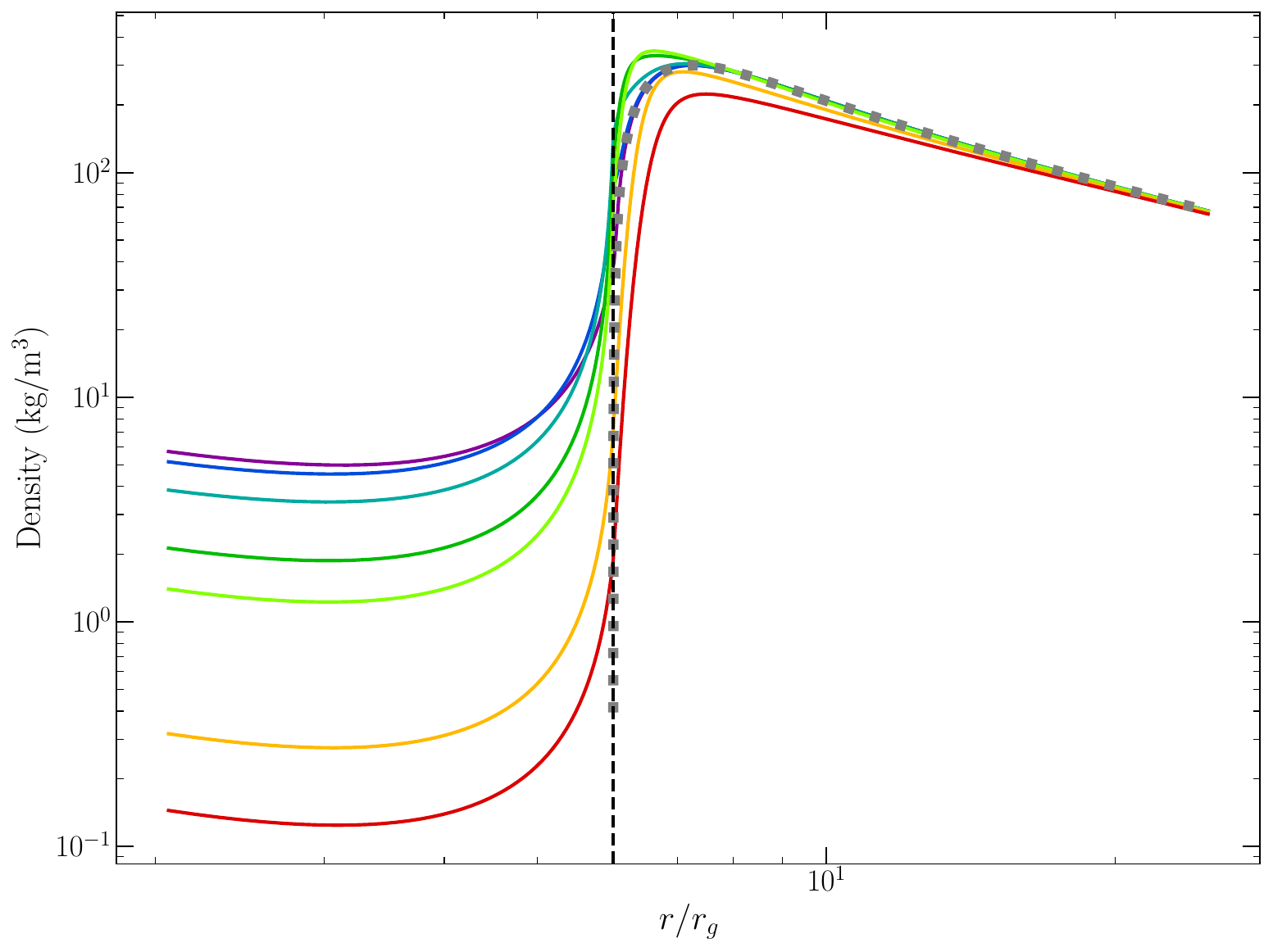}
\includegraphics[width=0.45\textwidth]{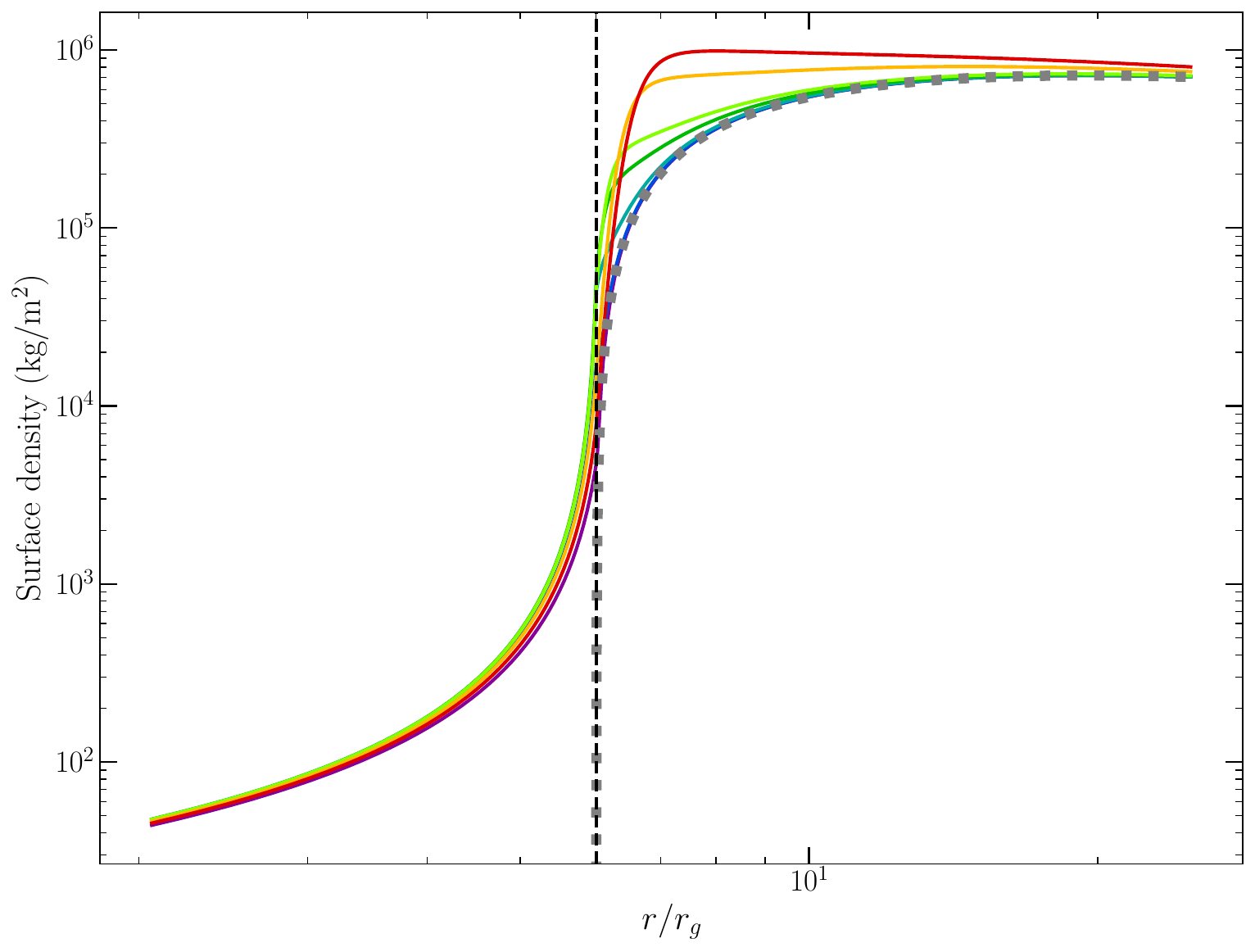}
\includegraphics[width=0.45\textwidth]{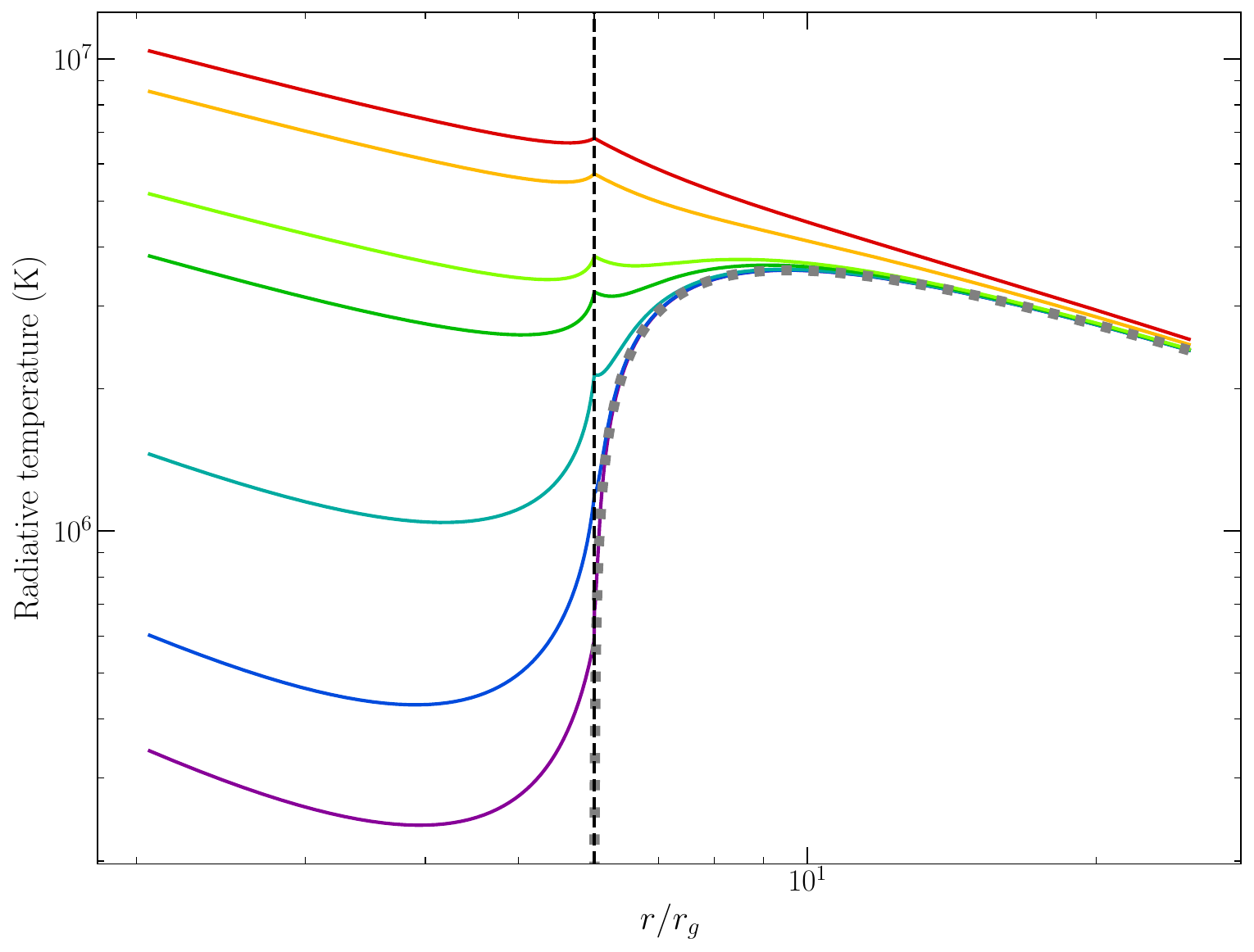}
\includegraphics[width=0.45\textwidth]{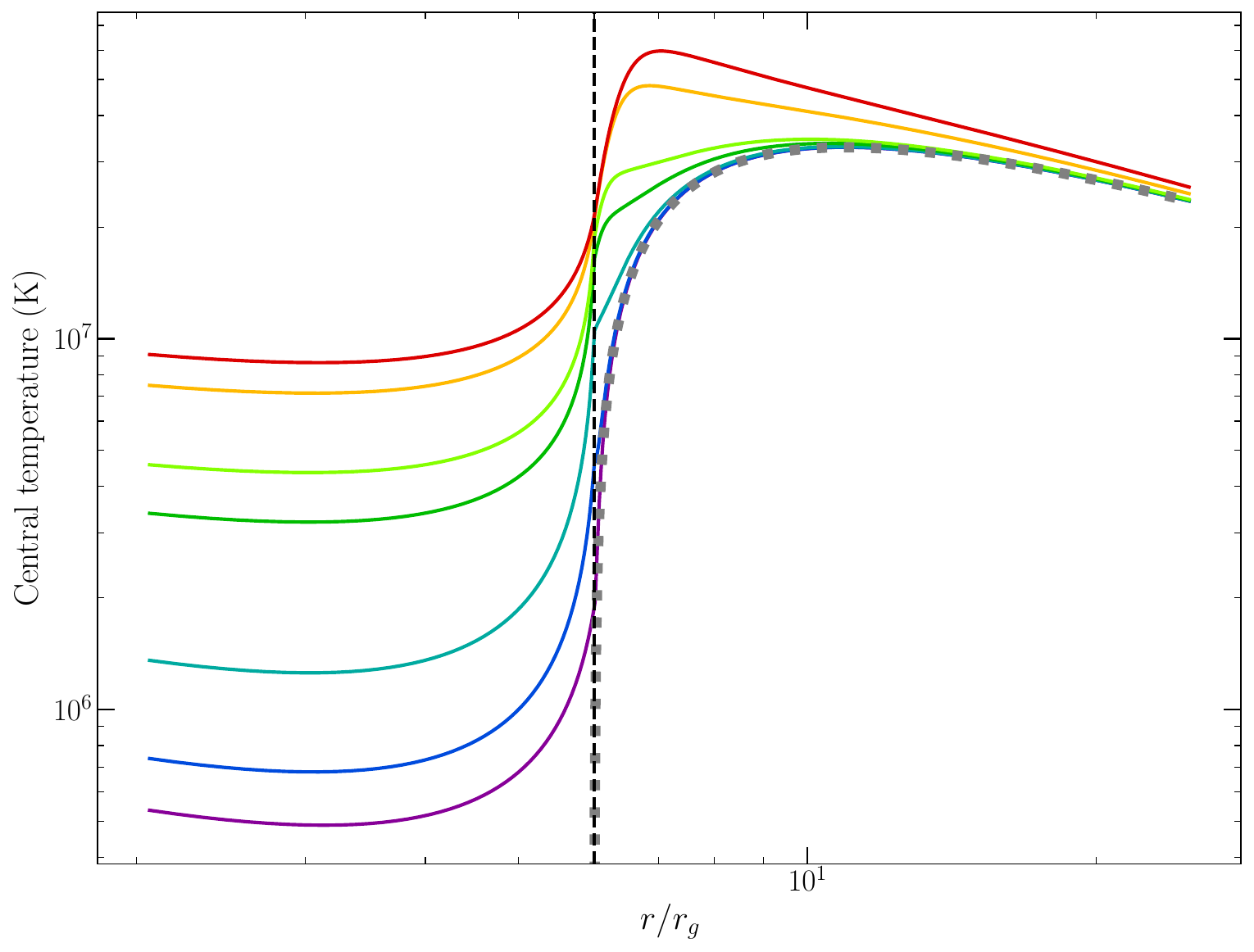}
\includegraphics[width=0.45\textwidth]{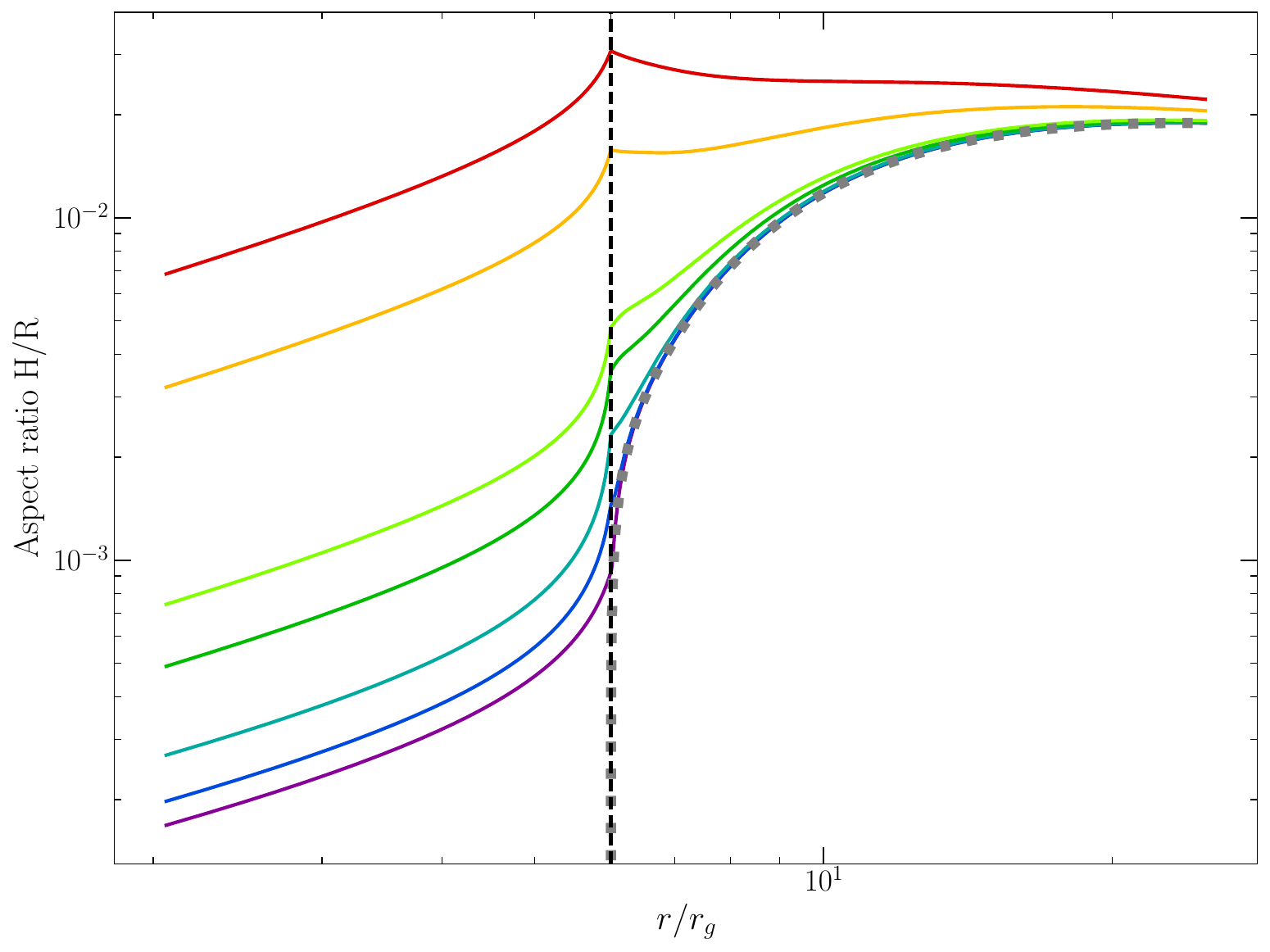}
\includegraphics[width=0.45\textwidth]{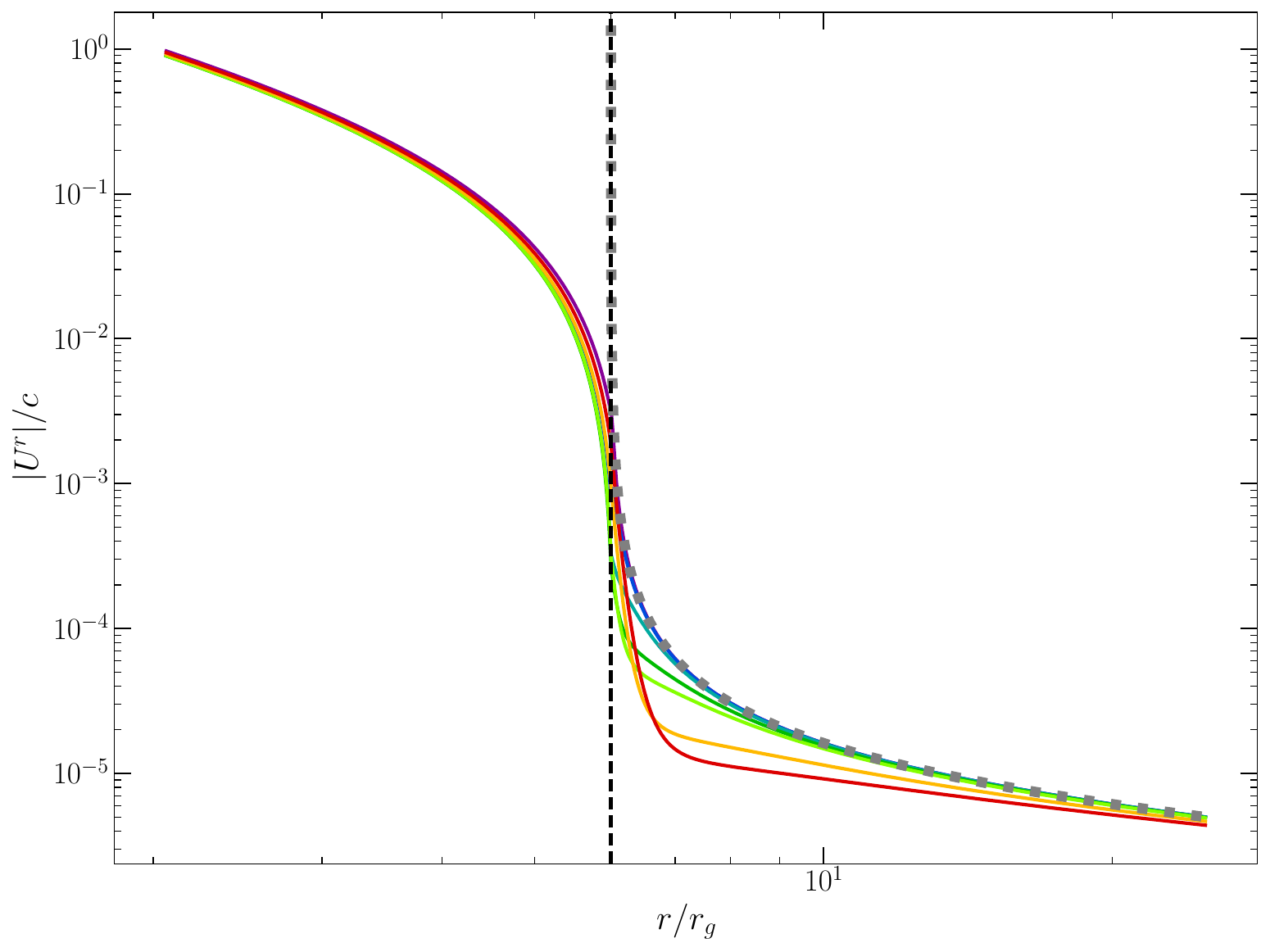}
\caption{Various different thermodynamic properties of a disc evolving about a Schwarzschild black hole with differing values of the dimensionless ISCO stress parameter $\delta_{\cal J}$. The ISCO radius is denoted by the vertical black dashed line. For reference, a vanishing ISCO stress solution is displayed by grey dots. The values of $\delta_{\cal J}$ used are $\log_{10} \delta_{\cal J} = -5, -4, -3, -2.3, -2, -1.3, -1$, where higher ISCO stresses can be identified by larger ISCO values of the radiative temperatures. The other parameters used in constructing this solution are $M_\bullet = 10 M_\odot, \dot M = 0.1 \dot M_{\rm edd}, \alpha = 0.1$.   }
\label{disc_soln}
\end{figure*}

\begin{figure*}    
    \centering
    \includegraphics[width=.75\linewidth]{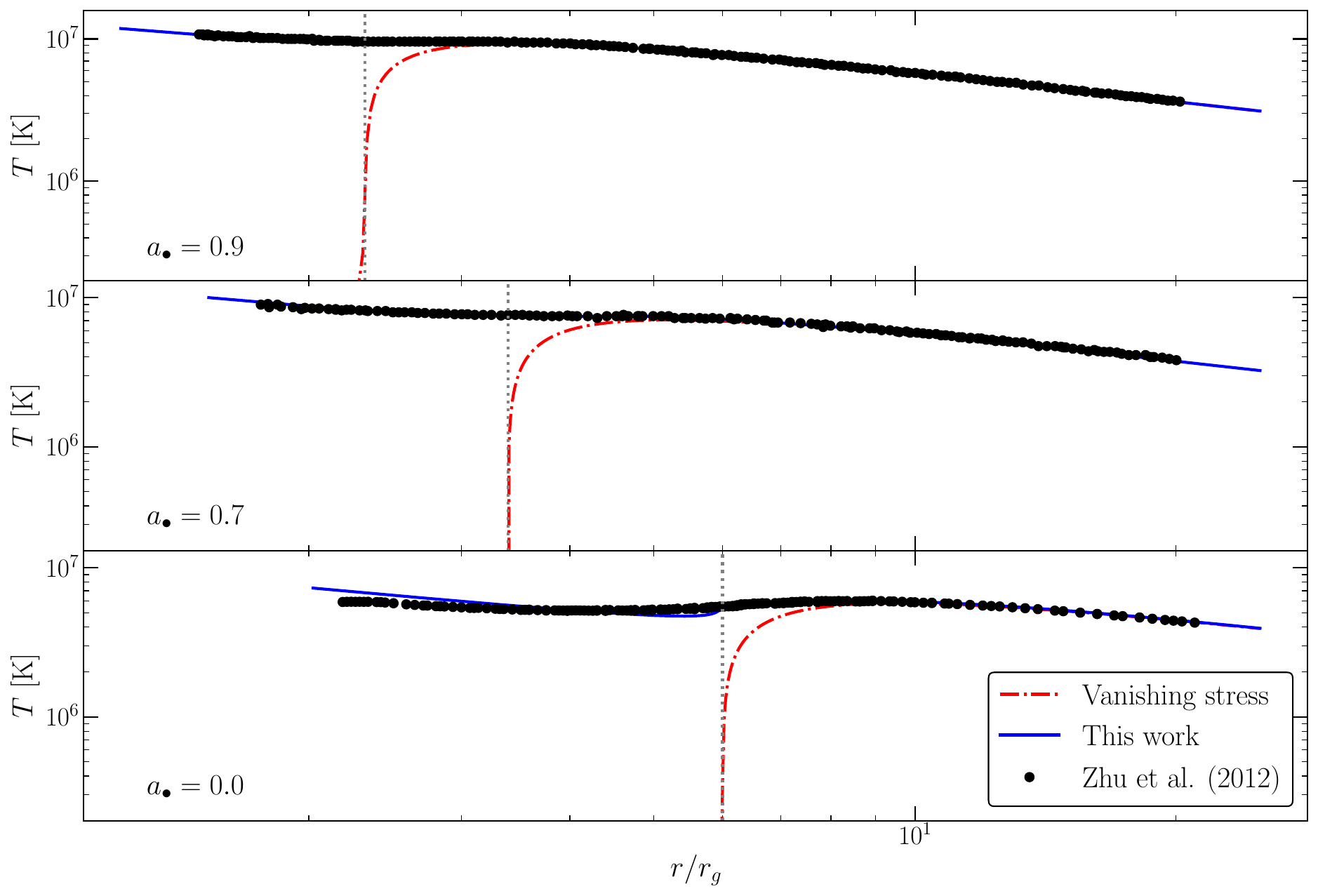}
    \caption{The radiative temperature profiles of \citealt{Zhu12} (black points; see text), compared to the models developed in this paper (blue solid curves) for three values of the black hole spin (displayed on plot). Also shown are vanishing ISCO stress models for comparison (red dashed curves). {The simple analytical models developed in this paper reproduce the gross behaviour of full GRMHD profiles.} }
    \label{fig:radtempcomp}
\end{figure*}

\subsubsection{Radial velocity profile}
We define a new velocity profile of the accretion flow solutions 
\beq
U^r(r) = U^r_{\rm NPT}(r) - \alpha^{1/2} c_{s, I} \, {\cal I}(r),
\eeq
where $U^r_{\rm NPT}$ denotes the classical solution \citep{NovikovThorne73, PageThorne74}, we define $U^r < 0$ for accretion {($U^r_{\rm NPT}$  is negative)}, and ${\cal I}(r)$ is an interpolation function, satisfying the constraints 
\beq
{\cal I}(r\to r_I) = 1, \quad {\cal I}(r \to \infty) = 0. 
\eeq 
The rational here of course is that the drift velocity is the asymptotically correct value for large radii, but that the non-zero directional bias in the fluctuations modifies the disc flow near to the ISCO. 

There are of course any number of functions which satisfy these asymptotic constraints. {There is an additional requirement on the interpolation function however, which helps narrow down the choice somewhat. We require that ${\cal I}(r)$ transitions from zero to one over a well defined scale length (which in the disc will be set by the turbulent eddy scale $\sim H$). We therefore turn to exponential functions, which have controllable length scales.   }

{We experimented numerically with simple exponential decays ${\cal I}(r) = \exp(-(r-r_I)/\Delta r)$, Gaussian decays ${\cal I}(r) = \exp(-(r-r_I)^2/\Delta r^2)$, and the following parameterisation}
\beq\label{interp}
{\cal I}(r) = \left(\tanh\left[ {\Delta r \over r - r_I }\right] \right)^4 .
\eeq
{All of which allow the radial scale over which the flow deviates from the classical description to be controlled through $\Delta r$} (the exponent 4 here is chosen so that ${\cal I}(r)$ goes to zero sufficiently quickly so as to not modify the disc thermodynamics on large scales). 

{We found no real qualitative difference between different choices of interpolation functions which shared a common tuneable length scale $\Delta r$. This is not particularly surprising, as each disc model has the same behaviour at the ISCO (as they share the same radial velocity), and the same behaviour at large radii \citep[given by the][disc solution]{NovikovThorne73}. As such, we simply display solutions with an interpolation function given by eq. (\ref{interp}), while noting that the properties of ${\cal I}(r)$ will be best constrained through future comparison to  numerical GRMHD experimentation. }

With ${\cal I}(r)$ specified, we can solve fully for the disc thermodynamic properties. One specifies the disc {and black hole} parameters $\dot M, M_\bullet, a_\bullet, \alpha$ and the ISCO stress $\delta_{\cal J}$, which specifies the radiative temperature profile $T_R(r)$. One then solves the boundary condition constraint for $c_{s, I}$, which gives the radial velocity profile through the above parameterisation. Mass conservation then gives the disc surface density, from which the central temperature, pressure, density and scale height of the disc can be calculated. 

Within the ISCO we employ the formalism of \cite{MummeryBalbus2023}, using the values of the various thermodynamic quantities at the ISCO as a boundary condition.

\subsection{Example solutions}
In this section we display the properties of some example solutions of the disc equations in this new framework. We take a moderately high ISCO stress (although we choose a value in the middle of the range found from GRMHD simulations), and choose other parameters suitable for a comparison to a typical X-ray binary. Explicitly we take $M_\bullet = 10 M_\odot$, $a_\bullet = 0$, $\alpha = 0.1$, and $\dot M = 0.1 \dot M_{\rm edd}$. We take a dimensionless ISCO stress parameter of $\delta_{\cal J} = 0.03$.  Thermodynamic disc properties for this solution, and a comparison to conventional models, are presented in Fig. \ref{Therm}.

We display with dashed curves the classical finite ISCO stress solutions, which display pronounced cusps at the ISCO (cf. Fig. \ref{cusps}), and by dotted curves the corresponding vanishing ISCO stress solution, which show an unphysical radial velocity dispersion $U^r \to \infty$ at the ISCO.  Finally, by solid curves we display the new solutions derived in this work, with a trans-ISCO velocity set by $\alpha^{1/2} c_s$, we see that the cuspy nature of the finite ISCO stress solutions has been removed, and the transition across the ISCO is significantly smoother. Each panel displays a different thermodynamic quantity, and the vertical axes of these plots are normalised by the value of the thermodynamic quantity at $10 r_g$, except for the radial velocity (lowest panel) which is plotted in units of the speed of light.  The new formalism put forward in this work nicely bisects the two traditional approaches.  

In Figure \ref{disc_soln} we plot various different thermodynamic properties of a disc evolving about a Schwarzschild black hole with differing values of the dimensionless ISCO stress parameter $\delta_{\cal J}$. The ISCO radius is denoted by the vertical black dashed line. For reference, a vanishing ISCO stress solution is displayed by grey dots. The values of $\delta_{\cal J}$ used are $\log_{10} \delta_{\cal J} = -5, -4, -3, -2.3, -2, -1.3, -1$, where higher ISCO stresses can be identified by larger ISCO values of the radiative temperatures (centre left panel). The other parameters used in constructing this solution are typical for galactic X-ray binaries $M_\bullet = 10 M_\odot, \dot M = 0.1 \dot M_{\rm edd}, \alpha = 0.1$.  Interestingly, in this new formalism, certain disc quantities show a much reduced dependence on the ISCO stress. This is particularly true for the disc surface density $\Sigma$, which shows barely any dependence on $\delta_{\cal J}$ despite it being varied by  4 orders of magnitude.   {This is because the surface density is set entirely by the radial velocity (through mass conservation $\dot M = 2\pi r \Sigma U^r$), and the trans-ISCO fluctuation velocity turns out to be only weakly dependent on the ISCO stress in this new formalism.  }

Other disc parameters remain more sensitively dependent on the local physics of the ISCO stress.   This is most notable for the radiative temperature of the flow $T_R$, which shows a dependence on ISCO stress within $r \lesssim 10r_g$, and is extremely sensitive to the ISCO stress at radii within the ISCO.  Similarly, the increased temperatures of these solutions leads to greater pressure support and notably different scale heights of each solution.   This filters through to a much reduced intra-ISCO density $(\rho)$ for larger ISCO stresses. This is of potential observational interest, as the disc density $\rho$ determines the ionisation fraction $(\xi)$ of the flow if the flow is illuminated by an incident X-ray flux $F_X$, $\xi \propto F_X / \rho$.  Lower densities from larger ISCO stresses will filter through to higher ionisation fractions, and correspondingly reduced iron line fluorescence.

It is clear from Fig. \ref{disc_soln} that some disc quantities still display a slight cusp at the ISCO, even within this new framework. This will remain an unavoidable effect of analytical models of trans-ISCO flows which involve the piecewise joining of intra- and extra-ISCO flows.  The present work minimises the presence of these kinks to as much of a degree as possible, and we do not believe that any remaining cusps will dramatically influence inferences from the fitting of observational data.   

\section{Comparison to GRMHD simulations}\label{grmhdcomp}

In this section we take our extended global
thin disc solutions and compare their thermodynamic profiles to those extracted from GRMHD simulations. A full comparison to dedicated numerical simulations is postponed to a future work, and we for now concentrate on potentially observable profiles extracted from previously published experiments. 

The two main potentially observable properties of a thin disc are the radial dependence of the “effective” (radiative) temperature profile, and the density of the flow. The radiative temperature is the key parameter of interest for so-called continuum fitting modelling of Galactic X-ray binaries \citep[see e.g.,][]{McClintock14}, as it directly determines the locally liberated flux in the fluids rest frame. As we discussed earlier, the density profile of a flow sets the ionisation fraction of the material when illuminated by an external X-ray flux $F_X$, with ionisation fraction $\xi \propto F_X / \rho$, and is therefore of direct interest to iron line studies \citep[see e.g.,][]{Reynolds13}. 

We first compare the radiative temperature profile of our model to those published in \cite{Zhu12}. \cite{Zhu12} extracted a radiative temperature from the local cooling rate computed in the GRMHD simulations run by \cite{Penna10}, and used them to examine some effects of the (neglect of the) plunging region on continuum fitting spin measurements. The radiative temperature profiles of \cite{Zhu12} are displayed in Figure \ref{fig:radtempcomp} by black dots, for three different black hole spins $a_\bullet =
0$ (lower panel), $a_\bullet = 0.7$ (middle panel) and $a_\bullet=0.9$ (upper panel). \cite{Zhu12} model an $M_\bullet = 10 M_\odot$ black hole accreting at roughly $\dot M \sim 0.1 \dot M_{\rm edd}$. We take these parameters as input to our analytical model. Various ``effective’’ $\alpha$ parameter was reported by \cite{Zhu12} and we take their values ($\alpha = 0.1$ for $a_\bullet = 0.7$ and $a_\bullet = 0.9$, $\alpha = 0.01$ for $a_\bullet=0$) for simplicity \citep[][their figure 6]{Zhu12}. We then only have the ISCO parameters to fit, namely $\delta_{\cal J}$.

We overplot in Figure \ref{fig:radtempcomp} the vanishing ISCO stress radiative temperature curve (red dashed curves), which are forced to zero at the ISCO contrary to the simulation results, and in blue (solid curves) the model developed in this paper. {We determine the appropriate value of the ISCO stress by minimising the loss function ${\cal L} = \sum_i \left(T_{\rm sim}(r_i) - T_{\rm disc}(r_i; \Delta r, \delta_{\cal J})\right)^2/T_{\rm sim}(r_i)^2$, where $T_{\rm sim}$ and $T_{\rm disc}$ are the simulation and analytical effective temperatures respectively.} The ISCO stress parameters are $\delta_{\cal J} = 0.0055$ for $a_\bullet = 0.9$, $\delta_{\cal J} = 0.0085$ for $a_\bullet = 0.7$ and  $\delta_{\cal J} = 0.007$ for $a_\bullet = 0$. We found no sensitivity to the fluctuation length scale parameter $\Delta r$, which we set to equal to the scale height of the disc in the solutions. We see that we recover the global properties of the GRMHD simulations {rather well} (note the inflection points in the radiation temperature profiles around the ISCO). This is an important result and motivates future development of extended continuum fitting models which include intra-ISCO emission. 

\begin{figure}    
    \centering
    \includegraphics[width=\linewidth]{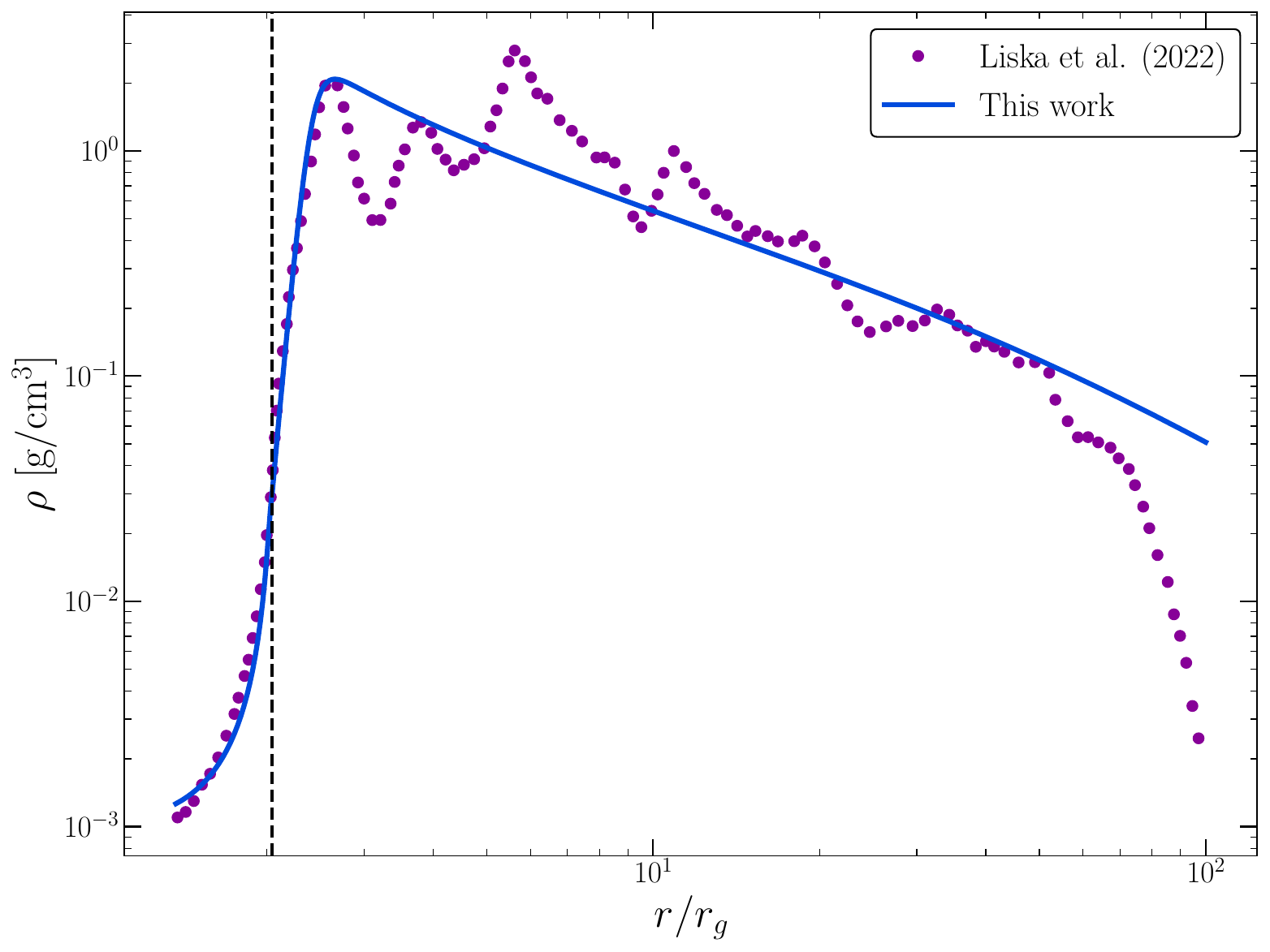}
    \caption{The density profile of the ``RADTOR'' simulation presented in \citealt{Liska22} (purple points), overplotted with the analytical model developed in this paper (blue curve). We find good agreement across the ISCO plunge, and also at large radii. }
    \label{fig:density-comp}
\end{figure}

Simulating the density profiles of thin GRMHD accretion flows with radiative transport effects included has only recently become computationally feasible \citep[e.g.,][]{Liska22, White23}. We extract the density profile from a SANE (i.e., ``standard and normal evolution''; the low magnetic field limit relevant for comparing to thin discs) simulation run by \cite{Liska22}, which was run for a spin $a_\bullet = 0.9375$ black hole, with mass $M_\bullet = 10 M_\odot$ \citep[][refer to this simulation as RADTOR in their paper]{Liska22}. The accretion rate in this simulation was set to be $\dot M \sim 0.35 \dot M_{\rm edd}$, and \cite{Liska22} extract an ``effective’’ $\alpha$ parameter which depended on radius but was roughly equal to $\alpha \sim 0.03$ a value we take in this work. The density profile extracted from \cite{Liska22} is plotted in Figure \ref{fig:density-comp}. 

Again, as the physical parameters of the main body of the disc are prespecified, we only fit  the intra-ISCO parameters of our model to the data. {The fit is performed by minimising the loss function ${\cal L} = \sum_i \left(\rho_{\rm sim}(r_i) - \rho_{\rm disc}(r_i; \Delta r, \delta_{\cal J})\right)^2/\rho_{\rm sim}(r_i)^2$, where $\rho_{\rm sim}$ and $\rho_{\rm disc}$ are the simulation and analytical density respectively. This loss function is dependent on the parameters $\Delta r$ and $\delta_{\cal J}$ only. The fit was performed only for $r \leq 50r_g$. } We find that the parameters $\delta_{\cal J} = 0.01$, $\Delta r / r_I = 0.03$ fits the global density profile well.  The deviation at large radii ($ r \gtrsim 75 r_g$) is likely a result of the finite mass reservoir in the GRMHD simulation. 

It is too early to say whether the fact that $\delta_{\cal J} \sim 10^{-2}$ best describes two different simulations which measure different disc quantities represents an interesting result, or is simply coincidence. 

As we move into the era where radiative GRMHD simulations of thin black hole discs become computationally feasible, we expect to perform many future tests of the models developed here. 

\section{Discussion}\label{general_discussion}
{In this paper we have identified a potential issue with classical modelling of black hole accretion flows -- by neglecting the finite turbulent eddy size of discs, classical models do not capture a directional bias in the turbulent velocity field induced by the ISCO, which acts much like a one-way gate for fluid elements. Ultimately, this bias has the dynamical effect that accretion flows typically cross the ISCO at significantly enhanced velocities (compared to the predictions of classical models).   }

{In aiming to elucidate this effect in a controlled manner, we have examined the properties of a set of random walk models. Random walk models are flexible, in that they allow both finite perturbations and global diffusive evolution to be examined simultaneously. They remain, of course, a simplification of the true dynamics of a turbulent MHD flow. In particular, the approximation that each fluid element behaves independently of all the others is an over simplification of the complex non-linear physics of a fluid. As these non-linear effects most likely mean that any one choice for a random walk velocity distribution (e.g., eq. \ref{fv_exp}) will be inaccurate, it is important that we have found that the qualitative results of the random  walk models are insensitive to this assumption. We also stress that the gross behaviour we are discussing can be intuited on purely physical grounds (see section \ref{discussion}).   }

{As the velocity profile of a black hole accretion flow will deviate from the classical $\alpha$-model predictions at short length scales from the ISCO \citep[e.g.][]{NovikovThorne73}, and the trans-ISCO velocity will be of order $\sim \alpha^{1/2} c_s$, we have introduced an interpolation function ${\cal I}(r)$ into our thin disc model radial velocity. This interpolation function must satisfy ${\cal I}(r\to \infty) \to 0$, and ${\cal I}(r\to r_I) \to 1$, with the transition expected to occur over a length scale of order one ISCO radius out from the ISCO, but is otherwise not currently well understood. Considering a class of exponential interpolation functions with tuneable transition scales, we did not find much sensitivity of the disc thermodynamic properties to the precise functional form chosen.   }

{Fortunately, when confronted with numerical data (e.g., Figure \ref{fig:density-comp}), the tuneable parameters in the interpolation functions can be constrained (and the new model fits well). As we move into the era where radiative GRMHD simulations of thin black hole discs become computationally feasible, we anticipate that the properties of the radial velocity transition will be well constrained by simulations, which are better suited to this task than analytical derivation.  }

{The fact that these extended models reproduce properties of full GRMHD simulations (Figs. \ref{fig:radtempcomp}, \ref{fig:density-comp}) is extremely promising. We stress that the modifications and extensions presented in this work are exclusive to the inner disc regime, and we expect classical $\alpha$-modelling to remain a good description of accretion flows at large radii. This can most clearly be seen in the reproduction of the large radius GRMHD simulation density profile in Figure \ref{fig:density-comp}. The good model fit extends well into the classical regime, where the modifications put forward in this paper have no effect.   }

{Where classical $\alpha$-modelling is clearly breaking down however is at small radial scales, typically at $\sim 1-2$ ISCO radii from the black hole. This can most clearly be seen in the cusps produced by classical models when ISCO stress parameters typically found in numerical simulations are substituted (Fig. \ref{cusps}, \ref{Therm}); these cusps are never themselves reproduced in simulations. While the new model derived in this paper smooths out these cusps to a significant degree (e.g., Figs. \ref{Therm}, \ref{disc_soln}), there are still some small scale cusps which remain at precisely the ISCO. This is an artifact which results from the piecewise joining of two distinct accretion solutions together, and we do not expect this behaviour to be physical. It seems likely to us that in a real fluid these cusps would be smoothed out, as is seen in GRMHD simulations (Fig. \ref{fig:radtempcomp}).  }

{In addition, the classical ``zero stress'' boundary condition imposed on $\alpha$-models clearly does not reproduce the properties of simulated accretion flows at small radii (Fig. \ref{fig:radtempcomp}). The ISCO stress parameters we infer from fitting to GRMHD data are at the scale $\delta_{\cal J} \sim 0.01$, which is at the smaller end of the range estimated from previous simulations \citep[for example][found $\delta_{\cal J}\sim 0.1$, while \citealt{Penna10} found $\delta_{\cal J} \sim 0.02$]{Noble10}. This is, however, certainly sufficiently large to show substantial deviations from disc profiles which assume zero stress (Fig. \ref{fig:radtempcomp}).   }

{The stress at the ISCO is fundamentally magnetic in origin \citep[e.g.,][]{Gammie99, AgolKrolik00}, and likely increases sharply with magnetisation \citep[e.g., the][model of the ISCO stress]{Gammie99}. As these new solutions reproduce properties of GRMHD simulations at a fraction of the computational cost, it is likely that we are heading into a future where observational constraints (like those inferred from continuum fitting and iron line modelling) will be able to place constraints on $\delta_{\cal J}$, the value of which has been a long controversial theoretical question.     }

\section{Conclusions }\label{conc}
In this paper we have examined the near-ISCO behaviour of thin black hole accretion flows, with a particular focus on the appropriateness of the {classical treatment of the mean fluid flow} in this limit. We have argued that because a turbulent flow (like MRI driven accretion) will display macroscopic {(of order the disc scale height)} perturbations in its velocity field, {the classical} description {(which does not distinguish scales above or below the turbulent eddy scale)} becomes an increasingly poor model as absorbing boundaries, such as the ISCO, are approached. Physically, this argument stems from the fact that in a turbulent flow large velocity fluctuations can carry a fluid element over the ISCO from a finite distance away, from which it will not return, a process without analogy in a {classical disc model}. This introduces a non-zero directional bias into the velocity fluctuations in the near-ISCO disc, a property which {is ordinarily ignored}. 

To examine the effects of this non-zero directional bias in the velocity fluctuations of accreting flows, we have examined the properties of some random walk models.  Random walk models are more mathematically flexible than purely viscous systems, and are well suited to modelling physical systems with both a global diffusive character (like an accretion flow at large radii), but also with large amplitude velocity fluctuations {(relevant for an accretion flow at small radii)}. It transpires that in the astrophysically relevant limit where the mean drift velocity of a fluid element $\epsilon$ is much smaller than the typical turbulent velocity fluctuations $V$ (in thin discs this ratio is of order $\epsilon / V \sim \alpha^{1/2} (H/R)\ll 1$) the typical velocity with which a fluid element crosses an absorbing boundary is of order the fluctuation scale $\sim {\cal O}(V) \gg \epsilon$. 

This increased radial velocity modifies the local thermodynamic quantities of this disc on either side of the ISCO. A practical application of this work is that it removes (as far as possible) cusps at the ISCO which are present in previous models of finite ISCO stress discs (e.g., Figs. \ref{cusps}, \ref{Therm}). This will be of practical importance when it comes to fitting analytical intra-ISCO models to data \citep[e.g.,][]{Reynolds97, Zhu12, Wilkins20} as such discontinuities might drive the overall fitting procedure. 

In this framework thin disc accretion around black holes is comprised of three fundamental regimes.  At large radii $r/r_I \gg 1$ accretion is diffusive (or ``viscous'') dominated, and standard models work well. At an order unity distance from the ISCO $r/r_I \sim 2$ the flow transitions to a fluctuation dominated state, and the fluid begins to ``learn'' about the ISCO. Within the ISCO $r/r_I < 1$ the flow transitions to near geodesic motion, and is gravitationally dominated. The analytical models we describe in this paper smoothly transition between the three regions. 

Finally, we have demonstrate that these new models are in good accord with the outputs of GRMHD simulations of thin discs. In Figure \ref{fig:radtempcomp} we demonstrate that the radiative temperature of these new solutions is in good accord with the results of \cite{Zhu12}, who computed the locally liberated flux from the GRMHD simulations of \cite{Penna10}. This local radiative temperature is the chief physical parameter which determines the thermal X-ray emission observed in X-ray binary soft states, and can therefore be directly probed with observations \citep[e.g.][]{McClintock14}. 

Similarly, we reproduce (Fig. \ref{fig:density-comp}) the density profile of the thin disc weak magnetic field simulation of \cite{Liska22}. Density profiles of discs can in principle be probed by the iron line fitting technique \citep{Reynolds13}, to which the plunging region provides a non-negligible contribution \citep{Reynolds97, Wilkins20}. It is our intention to use the models developed here and in \cite{MummeryBalbus2023} to extend ``continuum fitting'' and iron-line analysis procedures, to include emission and material inside the ISCO.

\section*{Acknowledgements}   
{The authors would like to thank the reviewer for a detailed report which improved the presentation of the results.} This work was supported by a Leverhulme Trust International Professorship grant [number LIP-202-014]. For the purpose of Open Access, the authors have applied a CC BY public copyright licence to any Author Accepted Manuscript version arising from this submission.   This work is partially supported by the Hintze Family Charitable Trust and STFC grant ST/S000488/1.

\section*{Data availability} 
No observational data was used in producing this manuscript. {The numerical random walk data will be shared upon request to the corresponding author. }

\bibliographystyle{mnras}
\bibliography{andy}

\appendix
\section{Explicit integral solutions for the exponential random walk}\label{appA}
In this appendix we list the explicit solutions for the integral definitions of the parameter $b$, the boundary crossing velocity $\left\langle v_b \right\rangle$, and the more general expression for the crossing velocity as a function of position. 
\subsection{The $b$-parameter }
The parameter $b$ is the solution of 
\begin{multline}
b \left(1 - {1 \over 2 \beta } \int\limits_{0}^{\infty} \exp\left(- {|\chi - x'| \over \beta} - {x' \over D}\right) \, {\rm d}x'\right) \\ = -1 + {1 \over 2 \beta } \int\limits_{0}^{\infty} \exp\left(- {|\chi - x'| \over \beta} \right) \, {\rm d}x' .
\end{multline}
or explicitly 
\begin{align}
b &= {-1 + I_1 \over 1 - I_2} , \\
I_1 &= {e^{-\chi/\beta} \over 2 \beta} \int_0^\chi e^{x/\beta} \, {\rm d}x + {e^{\chi/\beta} \over 2 \beta} \int_\chi^\infty e^{-x/\beta} \, {\rm d}x , \\
I_2 &= {e^{-\chi/\beta} \over 2 \beta} \int_0^\chi e^{\gamma x} \, {\rm d}x + {e^{\chi/\beta} \over 2 \beta} \int_\chi^\infty e^{-\gamma x} \, {\rm d}x , \\
{\rm where} \nonumber \\
\gamma &\equiv {1\over \beta}-{1\over D} .
\end{align}
Then 
\begin{align}
    I_1 &= 1 - {1\over 2} e^{-{\chi/\beta}} , \\ 
    I_2 &= {e^{-\chi/\beta} \over 2\beta \gamma} \left[ e^{\gamma \chi} - 1\right] + {e^{\chi/\beta -\chi \gamma} \over 2\beta \gamma } ,
\end{align}
and the parameter $b$ is determined. 
\subsection{The boundary crossing velocity} 
The boundary crossing velocity is given by the solution of the integral 
\begin{multline}
\left\langle v_b \right\rangle = {\int\limits_{-\infty}^0 v (bD -v\Delta t - bD \exp\left({v\Delta t \over D}\right)) \exp\left(-{|v+\epsilon|\over V}\right)   \, {\rm d}v \over   \int\limits_{-\infty}^0  (bD -v\Delta t - bD \exp\left({v\Delta t \over D}\right))  \exp\left(-{|v+\epsilon|\over V}\right) \, {\rm d} v} ,
\end{multline}
or explicitly 
\begin{align}
\left\langle v_b \right\rangle &= {I_1 + I_2 + I_3 \over I_4 + I_5 + I_6} , \\ 
I_1 &= - V^2 b D \left({2\epsilon \over V} + \exp\left(-{\epsilon \over V}\right) \right)   , \\
I_2 &= - V^3 \Delta t \left(2 \left({\epsilon \over V}\right) ^2 - 2 \exp\left(-{\epsilon \over V}\right) + 4\right), \\
I_3 &= {-V^2 b D \exp\left({\epsilon \Delta t \over D}\right) \over (V^2 \Delta t ^2 / D^2 -1)^2} \Bigg[ {2\epsilon \over V} \left({V^2 \Delta t ^2 \over D^2}-1\right) + { 4 V \Delta t \over D} \nonumber \\ & - \exp\left(-{\epsilon \over V} \left( {V\Delta t \over D} - 1\right)\right) \Bigg\{{2V \Delta t \over D} - 1 + {V^2 \Delta t ^2 \over D^2} \Bigg\} \Bigg] , \\
I_4 &= VbD\left(2 -\exp\left(-{\epsilon \over V}\right)\right) , \\
I_5 &= V^2 \Delta t \left({2\epsilon \over V} + \exp\left(-{\epsilon \over V}\right) \right) , \\
I_6 &= -{ VbD \exp\left(-{\epsilon \over V} \left(1 + {V\Delta t \over D}\right)\right) \over (V\Delta t / D)^2-1} \Bigg\{\left[1 + {V \Delta t \over D}\right] \exp\left({\epsilon \Delta t \over D}\right) \nonumber  \\ & -2\exp\left({\epsilon \over V}\right)\Bigg\}.
\end{align}
with small $\epsilon$ expansion (where we have used the following small $\epsilon$ results $D \to V^2 \Delta t / \epsilon$, and $b \to 1$)
\beq
\left\langle v_b \right\rangle = -{5\over 2 } V + \epsilon - {\cal O}(\epsilon^2) .
\eeq
\section{Robust numerical algorithms for solving transcendental equations }\label{appB}
For both the simple $\delta$-function jump distribution and the exponential jump distribution, the exact solutions of $p(x)$ require the solutions of a transcendental equation to be found. Simple gradient descent approaches do not always work robustly for these particular equations, as both have formal solutions where a parameter runs away to $ \pm \infty$, and so the ``wrong'' solution may be found. We present here a different approach to solving these equations which does not suffer from these difficulties.  

\subsection{Exponential random walk}
We wish to solve the transcendental equation 
\beq
\left({\beta \over D}\right)^2 = 1 - \exp\left(-{\chi \over D}\right),
\eeq
which has unique real solution $D>0$ provided $\chi, \beta > 0$. A simple numerical algorithm for solving this equation is based upon well known results from complex analysis. Start by defining  
\beq
F(z; \chi, \beta) = 1 - (\beta z)^2 - \exp(-\chi z) .
\eeq
If we are able to find the root $z_\star$ of $F(z_\star; \chi, \beta) = 0$, then we have found our required solution $D = 1/z_\star$.  If we take $z$ to be a complex variable, then the following fact is a direct result of Cauchy's residue theorem 
\beq
\oint_{\cal C} {z - z_\star \over F(z; \chi, \beta) } \, {\rm d} z = 0 ,
\eeq
which once rearranged gives 
\beq
D = {1 \over z_\star} = \left[ \oint_{\cal C} {1 \over F(z; \chi, \beta) } \, {\rm d} z \right] \Bigg/ \left[ \oint_{\cal C} {z \over F(z; \chi, \beta) } \, {\rm d} z \right] ,
\eeq
provided we carefully choose some contour ${\cal C}$ enclosing $z_\star$ and no other roots of $F$. In particular, it is important to avoid the trivial root at $z = 0$, which is never the physically required root. By rearranging the definition of $z_\star$ to 
\beq
z_\star = {\sqrt{1 - \exp(-\chi z_\star)} \over \beta} ,
\eeq
we see that $z_\star$ is bounded by 
\beq
{\chi\over \beta^2 } < z_\star < {1 \over \beta} . 
\eeq
The following circular contour always encloses $z_\star$ and only $z_\star$. The circular contour has centre at 
\beq
z_0 = \min(\chi/\beta^2, 1/\beta),
\eeq
with radius 
\beq
r = {1\over 2} z_0 .
\eeq
Substituting 
\beq
z(\phi) = z_0\left(1 + {1\over2} e^{i \phi}\right) , \quad \phi : 0\to2\pi ,
\eeq
into the above equation gives a simple numerical integral for $D$. 
\subsection{Simple jump distribution}
For the simpler jump distribution we need to solve the numerical equation 
\beq
2 = \exp(-\chi/D_2) + \exp(\beta/D_2) .
\eeq
We again define 
\beq
F(z; \chi, \beta) = 2 - \exp(-\chi z) - \exp(\beta z) ,
\eeq
and follow identical reasoning to before, so that 
\beq
D_2  = \left[ \oint_{\cal C} {1 \over F(z; \chi, \beta) } \, {\rm d} z \right] \Bigg/ \left[ \oint_{\cal C} {z \over F(z; \chi, \beta) } \, {\rm d} z \right] .
\eeq
The following circular contour always encloses $z_\star$ and only $z_\star$. The circular contour has centre at 
\beq
z_0 = {2 \chi - 2 \beta \over \chi^2 + \beta^2}
\eeq
with radius 
\beq
r = {1\over 2} z_0 .
\eeq
Substituting 
\beq
z(\phi) = z_0\left(1 + {1\over2} e^{i \phi}\right) , \quad \phi : 0\to2\pi ,
\eeq
into the above equation gives a simple numerical integral for $D_2$. 

\section{Energy conservation in thin discs}\label{appC}
We demonstrate in this section that the non-zero directional bias introduced into the radial velocity fluctuations does not modify the conservation of energy constraint in the disc, and therefore that this analysis does not modify the classical radiation temperature profiles \citep{NovikovThorne73, PageThorne74}. Start with the perfect fluid stress-energy tensor with radiative losses and a turbulent stress
\beq
T^{\mu\nu}= \left(\rho + {P + e \over c^2}\right) U^\mu U^\nu + \rho W^{\mu\nu} + P g^{\mu\nu} + {1\over c^2}(q^\mu U^\nu + q^\nu U^\mu),
\eeq
where $\rho$ is the rest mass density, $e$ the  energy density and $P$ the pressure of the fluid. The 4-velocity of the flow is $U^\mu$, while $U_\mu$ is its covariant counterpart. The {\it correlated} fluctuations in the flows 4-velocity produce a turbulent stress $W^{\mu\nu} \equiv \left\langle \delta U^\mu \delta U^\nu \right\rangle$.   The final pair of terms represent the energy-momentum flux carried out of the system by photons, where $q^\mu$ is the photon flux 4-vector. Energy-momentum conservation is expressed as $\nabla_\mu T^{\mu\nu} = 0$, or as will be more convenient for our purposes 
\beq
\nabla_\mu T^\mu_0 =0, \quad \nabla_\mu T^\mu_\phi = 0, 
\eeq
where in these expressions $\nabla_\mu$ is a covariant derivative with respect to Kerr metric coordinate $x^\mu$, and the left hand expression expresses energy conservation, while the right hand expression describes angular momentum conservation. Note that, for any mixed tensor $S^{\mu}_\nu$
\beq
\nabla_\mu S^\mu_\gamma = {1\over \sqrt{|g|}} \, \partial_\mu \left(\sqrt{|g|} S^\mu_\gamma\right)  - \Gamma^\lambda_{\mu \gamma} S^\mu_\lambda = 0, 
\eeq
where we have introduced the affine connection 
\beq
\Gamma^\alpha_{\beta \gamma}  \equiv {1\over 2} g^{\alpha \delta} \left( \partial_\beta g_{\gamma \delta} + \partial_\gamma g_{\beta \delta} - \partial_\delta g_{\beta \gamma} \right) . 
\eeq
For metrics which do not depend explicitly on coordinate $x^\gamma$ (the Kerr metric does not depend on $t$ and $\phi$) $\partial_\gamma g_{\mu\nu} = 0$, and so 
\beq\label{Ch2GammaFact}
\Gamma^\lambda_{\mu \gamma} = {1\over 2} g^{\lambda \delta} \left(\partial_\mu g_{\gamma \delta} - \partial_\delta g_{\mu\gamma} \right) ,
\eeq
meaning the combination 
\begin{multline}\label{Ch2GammaFact2}
\Gamma^\lambda_{\mu \gamma} S^\mu_\lambda = {1\over 2} g^{\lambda \alpha} \left(\partial_\mu g_{\gamma \alpha} - \partial_\alpha g_{\mu\gamma} \right)S^\mu_\lambda\\ = {1\over 2} \left(\partial_\mu g_{\gamma \alpha} - \partial_\alpha g_{\mu\gamma} \right)S^{\mu\alpha}  \equiv 0 ,
\end{multline}
vanishes for {\it any} symmetric tensor $S^{\mu\alpha}$, since the metric derivative are anti-symmetric in $\mu$ and $\alpha$, while $S^{\mu \alpha}$ is symmetric in these indices.   As a result of this identity, the conservation of disc angular momentum and energy become 
\beq
 {1\over \sqrt{|g|}} \, \partial_\mu \left(\sqrt{|g|} T^\mu_\phi\right) = 0, \quad  {1\over \sqrt{|g|}} \, \partial_\mu \left(\sqrt{|g|} T^\mu_0\right) = 0.
\eeq
Expanding, and using $\nabla_\mu g^{\mu\nu} \equiv 0$, and assuming that $\rho c^2 \gg P + e$, we are left with 
 \beq
 {1\over \sqrt{g}} \, {\partial \over \partial x^\mu} \Big[\sqrt{g}\rho\left(U^\mu U_0 + W^\mu_0\right) + \sqrt{g}(q^\mu U_0 + U^\mu q_0) \Big] = 0   ,
\eeq
which upon expanding (and neglecting the asymptotically small $q_0$ term) is 
\begin{multline}
U_0 \left[ {1 \over \sqrt{g}} \partial_\mu \left(\sqrt{g} \rho U^\mu\right) \right] + \rho U^r \partial_r U_0 \\ + {1 \over \sqrt{g}} \partial_\mu \left(\sqrt{g} \rho W^\mu_0 \right) + U_0 \partial_z q^z = 0 .
\end{multline}
The first term in square brackets in the above expression is just mass conservation within the disc, and is zero. Thus, energy conservation leads to 
\beq\label{econ}
\nabla_\mu \left(T^\mu_0\right) =  \rho U^r \partial_r U_0 + {1 \over \sqrt{g}} \partial_\mu \left(\sqrt{g} \rho W^\mu_0 \right) + U_0 \partial_z q^z = 0 .
\eeq 
Identical reasoning as the above produces a symmetric (with 0 replaced by $\phi$) equation of conservation of angular momentum 
\beq\label{angcon}
\nabla_\mu \left(T^\mu_\phi\right) =  \rho U^r \partial_r U_\phi + {1 \over \sqrt{g}} \partial_\mu \left(\sqrt{g} \rho W^\mu_\phi \right) + U_\phi \partial_z q^z = 0 . 
\eeq 
To derive the energy equation of the flow, take $U^0$  times the energy conservation equation and add it to $U^\phi$ times the angular momentum conservation equation. This procedure leaves 
\begin{multline}\label{Ch2econ3}
\rho U^r \left[ U^0 \partial_r U_0 + U^\phi \partial_r U_\phi \right] +  {U^0 \over \sqrt{g}} \partial_\mu \left(\sqrt{g} \rho W^\mu_0 \right) \\ + {U^\phi \over \sqrt{g}}  \partial_\mu \left(\sqrt{g} \rho W^\mu_\phi \right)   = -(U^0U_0 + U^\phi U_\phi )  \partial_z q^z ,
\end{multline}
which is the result used in the paper. 

\section{ Boundary condition for $c_{s, I}$ } \label{appD}
To derive the governing boundary condition expression for the ISCO speed of sound, one must solve the coupled equations 
\begin{align}
c_{s, I}^2 &\equiv {P_I \over \rho_I} , \label{D1}\\
H_I &= c_{s, I} \sqrt{r_I^3 \over 2 G M_\bullet} 
 ,\label{D2} \\
\rho_I &\equiv {\Sigma_I \over H_I}, \label{D3}\\
P_I &\equiv {\rho_I k T_{c, I} \over \mu m_p} + {3 \sigma T_{c, I}^4 \over 3 c} ,\label{D4} \\
T_{c, I}^4 &= \left({3\kappa \Sigma_I\over 8} \right) T_{R, I}^4 , \label{D5}\\
\Sigma_I &= {\dot M \over 2\pi r_I \alpha^{1/2} c_{s, I}} \label{D6}. 
\end{align}
These are, in order, the definition of the speed of sound, the solution of vertical hydrostatic equilibrium, the definition of the disc density and pressure, the approximate solution of radiative transfer in the disc atmosphere, and the conservation of mass in the disc.  Note that we have used the exact result $U_{\phi, I}^2 - a^2 c^2 (1 - U_{0, I}^2) = 2 GM_\bullet r_I$ in the equation of hydrostatic equilibrium \citep[proof in][]{MummeryBalbus2023}. 

Substituting \ref{D5} into \ref{D4}, before substituting \ref{D4} and \ref{D3} into \ref{D1} leaves to a simple expression for $c_{s, I}$ in terms of known quantities ($\alpha, \dot M, r_I$) and the quantities  $H_I$ and $\Sigma_I$. This final expression can then be expressed entirely in terms of $c_{s, I}$ and known quantities by the substitution of \ref{D2} and \ref{D6}. The final result is as shown in the main body of the paper, namely 
\begin{equation}
c_{s, I}^2 = {\sigma \kappa T_{R, I}^4  \over 2 c} \sqrt{r_I^3 \over 2 G M_\bullet} c_{s, I} + \left({3\kappa\dot M \over 16  \pi r_I \alpha^{1/2} c_{s, I}}\right)^{1/4} {k T_{R, I}\over \mu m_p} .
\end{equation}

\label{lastpage}
\end{document}